%% file: paper_triangulation_graphs_jstatphys.tex
\documentclass[smallextended]{svjour3}                    
\smartqed  
\journalname{J Stat Phys}
\usepackage{mathptmx}      

\usepackage{graphicx}  
\usepackage{bbm}       
\usepackage{amsmath}
\usepackage{amssymb}   

\usepackage{braket}
\usepackage{extarrows}
\usepackage{multirow}
\usepackage{bbm}
\usepackage{layouts}
\usepackage{color}

\newcommand{\erdoes}{Erd\"os-R\'{e}nyi }
\newcommand{\erdoesnospc}{Erd\"os-R\'{e}nyi}
\newcommand{\barabasi}{Barab\'{a}si-Albert }

\begin{document}

\title{Spectral properties of unimodular lattice triangulations}
\titlerunning{Spectral properties of unimodular lattice triangulations}
\author{Benedikt Kr\"uger \and Ella Schmidt \and Klaus Mecke}

\institute{B. Kr\"uger \at FAU Erlangen-Nuremberg, Institute for Theoretical Physics, Staudstr. 7, 91058 Erlangen, Germany \\
  Tel.: +49-9131-8528451 \\
  Fax: +49-9131-8528444\\
  \email{benedikt.krueger@fau.de}
}
\date{Received: date / Accepted: date}

\maketitle

\begin{abstract}
Random unimodular lattice triangulations have been recently used as an embedded random graph model, which exhibit a crossover behaviour between an ordered, large-world and a disordered, small-world behaviour. 
Using the ergodic Pachner flips that transform such triangulations into another and an energy functional that corresponds to the degree distribution variance, Markov chain Monte-Carlo simulations can be applied to study these graphs.  
Here, we consider the spectra of the adjacency and the Laplacian matrix as well as the algebraic connectivity and the spectral radius. Power law dependencies on the system size can clearly be identified and compared to analytical solutions for periodic ground states. 
For random triangulations we find a qualitative agreement of the spectral properties with well-known random graph models. 
In the microcanonical ensemble analytical approximations agree with  numerical simulations. In the canonical ensemble a crossover behaviour can be found for the algebraic connectivity and the spectral radius, thus combining large-world and small-world behavior in one model. 
The considered spectral properties can be applied to transport problems on triangulation graphs and the crossover behaviour allows a tuning of important transport quantities.


\end{abstract}

\input{introduction.tex}

\input{triangulations_graphs.tex}

\input{analytical_solution.tex}

\input{results.tex}
\input{results_random.tex}
\input{results_microcanonical.tex}
\input{results_canonical.tex}

\input{conclusions.tex}

\bibliographystyle{spmpsci}
\bibliography{literature_laplacian}

\end{document}

%% file: introduction.tex
\section{Introduction}\label{sec:introduction}
Many real world systems consist of a set of equivalent objects and the pairwise interaction among themselves, so that they can be described approximately in terms of graph theory \cite{Newman_2010,Costa_2011}.
Excluding all quantitative aspects of the interactions and taking only into account whether two objects interact or not, the considered objects are interpreted as the node of a graph, where edges exist between objects or nodes that do interact.
Examples for such systems are neural networks in biology, where dendrites are modelled as nodes and axons between two dendrites are modelled as edges; 
or the world-wide-web, where an edge between two website nodes exist if there is a link from one website to the other.

A major tool in studying the physical properties of such graphs is spectral graph theory, which examines the spectra of the adjacency and the Laplacian matrices associated with the graph and some special eigenvalues of those, see e.g.~\cite{Cvetkovic_2010}.
The spectra of graphs are studied e.g. for quantum percolation \cite{Evangelou_1983} and Anderson transition on Bethe lattices \cite{Mirlin_1991,Evangelou_1992,Evangelou_1992b} in terms of random matrices, as well as in biology \cite{Banerjee_2009} and chemistry \cite{Trinajstic_1991,Estrada_2000}, for a review see \cite{Mohar_1991} and the references therein.
Special eigenvalues as the algebraic connectivity, the smallest non-zero eigenvalue of the Laplacian matrix, are important for characterizing the topology of graphs \cite{Juvan_1993}, for transport and dynamics on networks \cite{Almendral_2007} and for optimization problems \cite{Mohar_1993,Varshney_2013}, see \cite{Mohar_1992,deAbreu_2007} for reviews.
Additionally, spectral graph theory is well-connected with the common mathematical theory of random matrices used in quantum physics \cite{Guhr_1998}.
An important application of the spectrum of the Laplacian matrix are (quantum) random walks on networks \cite{Bray_1988,Monasson_1999}, where the Laplacian matrix corresponds to the discretization of the  operator used in Poisson and Schr\"odinger equations.
In both cases the spectrum can be used for calculating return probabilities and inverse participation ratios.

For understanding and modelling the important features of real world networks artificially constructed graphs can be used, mainly random graphs \cite{Albert_2002}.
Common examples are the \erdoes random graph \cite{Erdoes_1959,Erdoes_1960,Erdoes_1961}, the Watts-Strogatz random graph \cite{Watts_1998} (and a slightly altered version denoted as Newman-Watts random graph \cite{Newman_1999}) and the \barabasi random graph \cite{Barabasi_1999}.
These random graphs were mainly constructed and optimised for creating models for real small-world and scale-free networks, but also their spectral properties arise much interest \cite{Muelken_2011,Almendral_2007}.

Recently, a new type of (embedded) random graph model was proposed by using canonical ensembles of unimodular lattice triangulations \cite{Krueger_2015}.
Random lattice triangulations are comparable with the usual random graph models; in the canonical ensemble a crossover behavior from ordered triangulations showing large-world to disordered triangulations exhibiting small-world and scale-free behavior can be found.
The triangulations used  are tessellations of the convex hull of a given point set into triangles that do not intersect \cite{DeLoera_2010}.
Triangulations have been used as random-graphs before \cite{Andrade_2005,Zhou_2005,Song_2012,Aste_2012,Kownacki_2004,Almeida_2013}, but only the topological degrees of freedom were used there.
In contrast to these models the actual coordinates of the nodes become important in lattice triangulations \cite{Krueger_2015}, which adds some (numerical) difficulties, but can be a concept useful in real networks where spatial coordinates of  nodes and the induced length of the edges become important.
The considered lattice triangulations are maximal planar graphs in the sense that no interior edge can be inserted without violating the planarity. 
Furthermore, one can always find an embedding of a planar graph with rational coordinates, by choosing first an arbitrary embedding with real coordinates and then wiggling the non-rational ones, since the rational numbers are dense in the reals.
By scaling one can find an embedding with integer coordinates, so lattice triangulations can be considered as the supergraphs of all planar graphs.

Triangulations in general are used in quantum geometry for describing curved space-times in the approaches of causal dynamical triangulations \cite{Ambjorn_2005} and in spin foams \cite{Rovelli_2007}. In topology and geometry one is interested in the number of triangulations of manifolds \cite{Sulanke_2009}.
They can also be used as a tool for studying foams \cite{Sullivan_1999,Oguey_2003,Aste_1995,Dubertret_1998b}, since the Voronoi tessellation is the dual of the Delaunay triangulation, which is a special triangulation of a point set.

In this paper we extend the results of \cite{Krueger_2015} by considering spectral properties of lattice triangulations in order to relate graphs to physical properties. 
We measure the spectra of the adjacency and the Laplacian matrix of random triangulations, which is the ensemble of all triangulations with constant weights, using Metropolis Monte Carlo simulations, and compare the results with common models of random graphs.
Defining an energy of a triangulation, which is well known in the literature, and that corresponds to the variance of the degree distribution (the histogram of the node degrees, which is the number of incident edges), we calculate the microcanonical and canonical expectation values of the algebraic connectivity and the spectral radius with Metropolis and Wang-Landau simulations.
For the maximal ordered triangulations, which are the ground state of the energy function used, analytical calculations are possible, as well as approximations for triangulations with low energies near the ground state.
As an application of our results, we calculate the inverse participation ratio of certain eigenstates and show that random lattice triangulations on average exhibit stronger localization than comparable random graphs.

%% file: triangulations_graphs.tex
\section{Triangulations and spectral graph theory}\label{sec:triangs_graphs}

\subsection{Unimodular lattice triangulations}\label{subsec:lattice_triangulations}

\begin{figure}
  \includegraphics[width=\columnwidth]{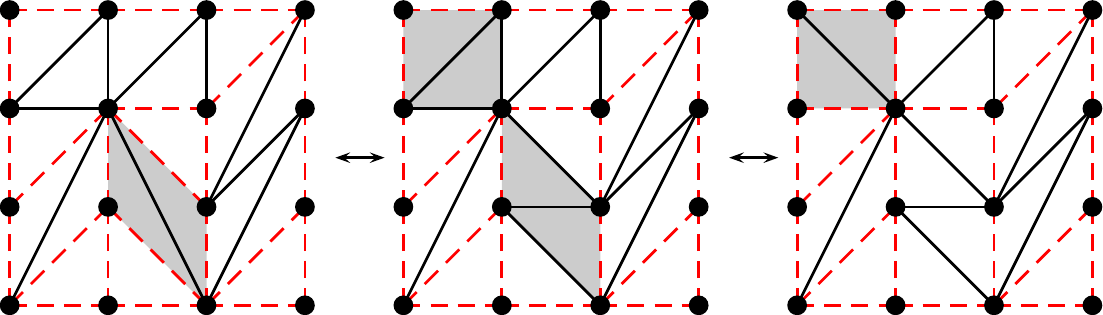}
  \caption{\label{fig:pachner_move} (Color online) Pachner moves in lattice triangulations. Every black, solid edge can be flipped into the other diagonal of the quadrangle consisting of the two triangles incident with this edge. The red, dashed edges cannot be flipped because they would lead to structures that are not triangulations. The flips associated with the grey quadrangles are executed from left to right and vice versa. Note that the flippability of an edge can change if neighboring edges are flipped.}
\end{figure}

\begin{figure}
  \includegraphics[width=0.3\columnwidth]{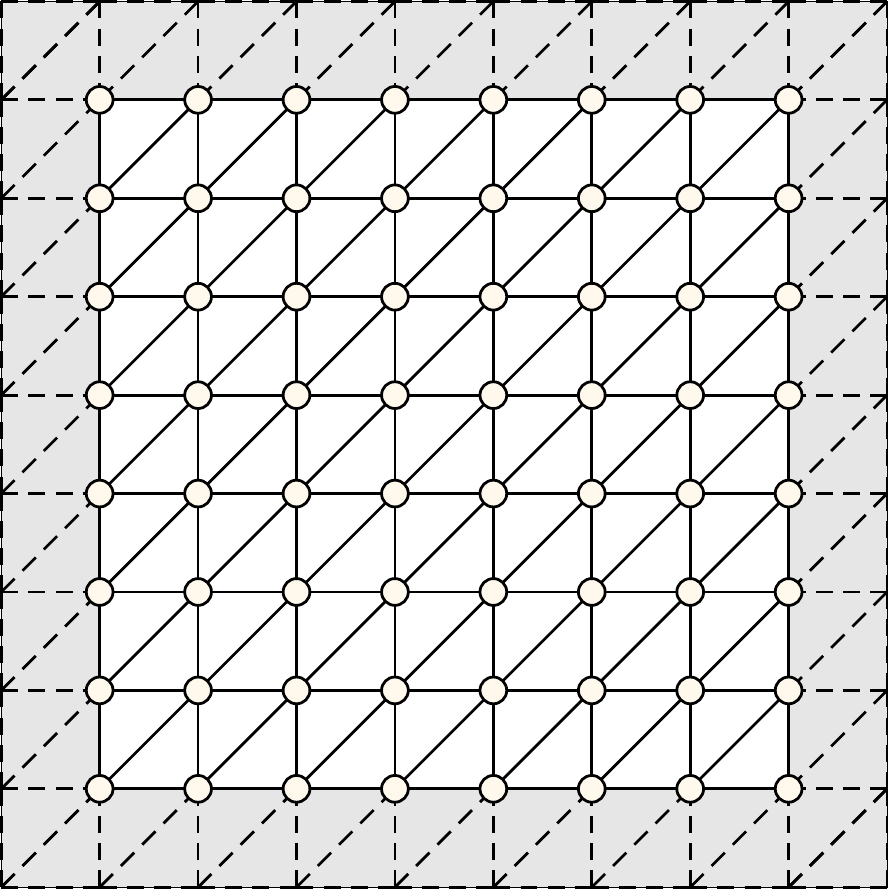}
  \hspace{0.025\columnwidth}
  \includegraphics[width=0.3\columnwidth]{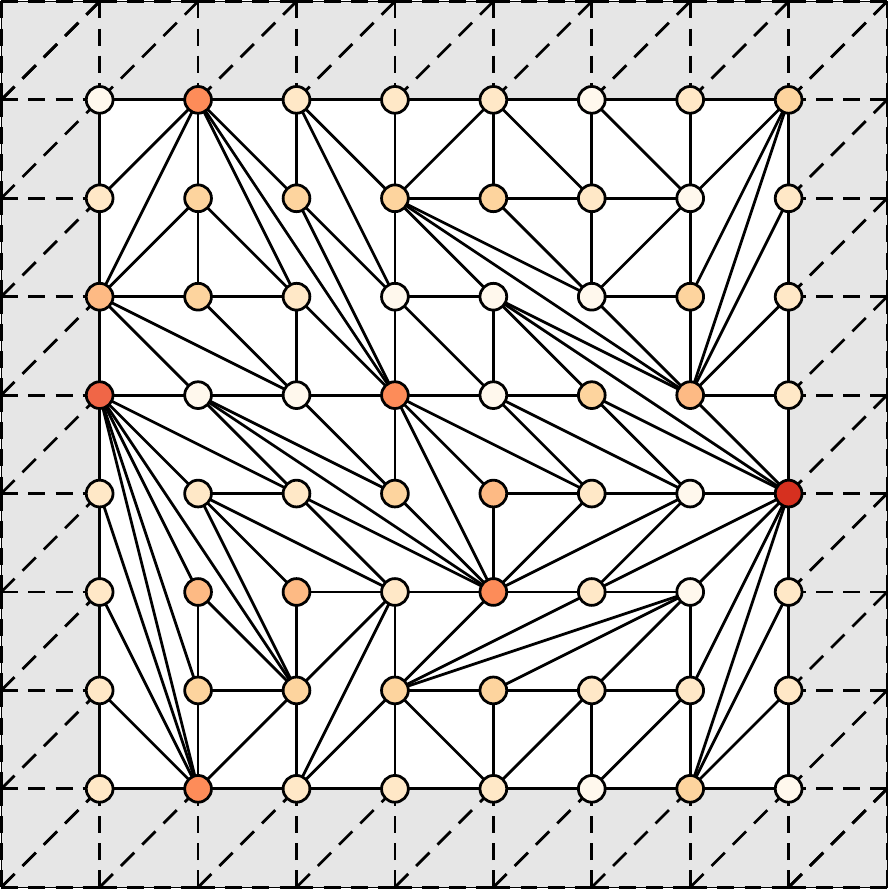}
  \hspace{0.025\columnwidth}
  \includegraphics[width=0.3\columnwidth]{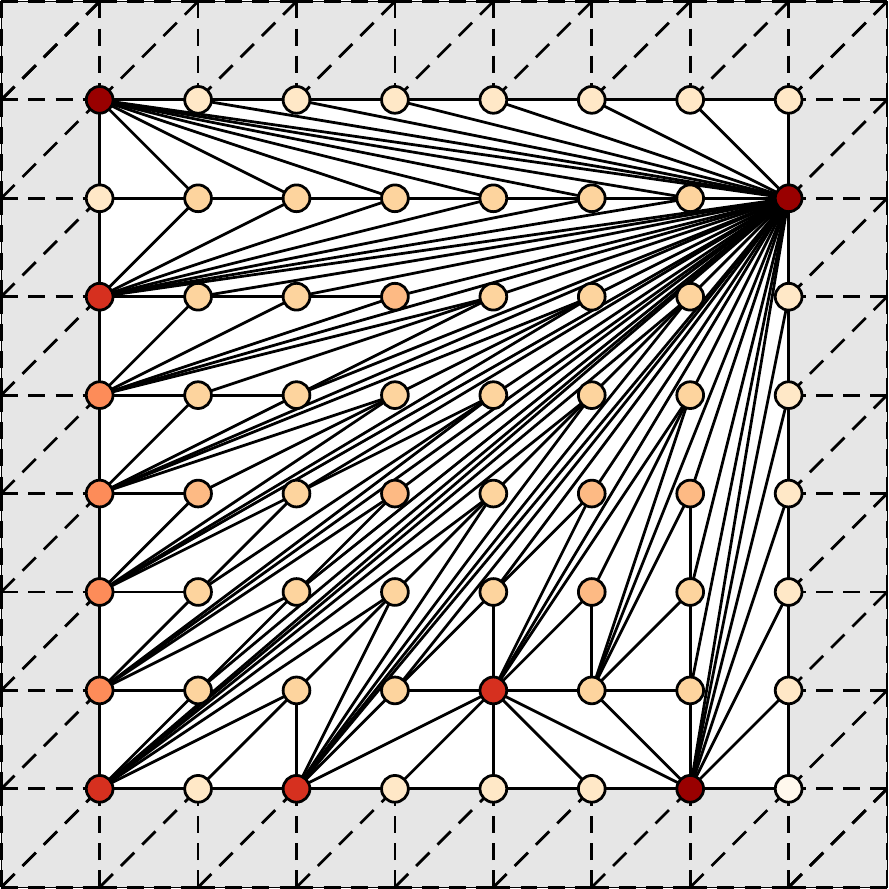}\\
  \makebox[0.3\columnwidth]{$\beta \rightarrow +\infty$}
  \hspace{0.025\columnwidth}
  \makebox[0.3\columnwidth]{$\beta = 0$}
  \hspace{0.025\columnwidth}
  \makebox[0.3\columnwidth]{$\beta \rightarrow -\infty$}
  \caption{\label{fig:triang_examples}(Color online) Examples of unimodular triangulations of a $8\times8$ lattice. The color of the vertices corresponds to the number of incident vertices, the dashed border is fixed and will not be flipped, but is considered for the calculation of the energy. From left to right: Maximal ordered triangulation, which is used as a reference state in the energy function \eqref{eq:energy} and therewith the ground state of this energy function ($\beta \rightarrow +\infty$), random triangulation at infinite temperature ($\beta = 0$), example of a maximal energy configuration ($\beta \rightarrow -\infty$).}
\end{figure}

In this section we follow the notations of \cite{DeLoera_2010,Krueger_2015}.
Let $\mathbf A = \{x_1, \dots x_K\}, x_i \in \mathbb R^2$ be a finite set of points in the plane and let $\mathrm{conv}(\mathbf A)$ be the convex hull of this point set~$\mathbf A$.
A triangulation $\mathcal T(\mathbf A)$ of the point set $\mathbf A$ is a tessellation of $\mathrm{conv}(\mathbf A)$ into triangles $\sigma_i \in \mathbf A^3$ generated by three points of $\mathbf A$, so that firstly the interiors $\sigma_i \setminus \partial \sigma_i$ and $\sigma_j \setminus \partial \sigma_j$ of two distinct triangles $\sigma_i \neq \sigma_j$ do not intersect and so that secondly each point is corner point of at least one triangle.
Sometimes the second defining property is dropped and triangulations that fulfil this property are called full.
In this paper we consider only full triangulations of two-dimensional integer $M\times N$-lattices
\begin{equation*}
  \mathbf A = \left\{ \begin{pmatrix} m \\ n \end{pmatrix} \mid m \in \mathbb Z_M := \{0, 1, \dots, M-1\}, n \in \mathbb Z_N \right\}.
\end{equation*}
These triangulations are unimodular, that is that all triangles have equal area $1/2$. 
Examples of such triangulations can be found in Fig.~\ref{fig:pachner_move} and Fig.~\ref{fig:triang_examples}.
Each triangulation is also a simplicial complex if one includes for each triangle $\sigma_i = \{x_{i_1}, x_{i_2}, x_{i_3}\} \in \mathcal T$ the 1-simplices (edges) $\{x_{i_1}, x_{i_2}\}$, $\{x_{i_1}, x_{i_3}\}$ and $\{x_{i_2}, x_{i_3}\}$ that consist of two of the triangle points, and the 0-simplices (vertices) $\{x_{i_1}\}$, $\{x_{i_2}\}$ and $\{x_{i_3}\}$ consisting of one point.

One special triangulation used throughout this paper is the maximal ordered triangulation.
  It is defined as the full triangulation of the $M\times N$ integer lattice where the vertex with coordinates $(i, j)$ is connected with at most six vertices with coordinates $(i, j \pm 1)$, $(i \pm 1, j)$ and $(i \pm 1, j \pm 1)$ whenever these coordinates are in $\mathbb Z_M$ or $\mathbb Z_N$ respectively.
Note that postulating connections to vertices with coordinates $(i \pm 1, j \pm 1)$ is a convention, one could also use connections to $(i \pm 1, j \mp 1)$ instead.
The maximal ordered lattice is often denoted as triangular lattice in the literature.

To determine the number of possible triangulations of an integer lattice is a non-trivial question.
There are exact enumeration results and analytical bounds \cite{Kaibel_2003} as well as numerical approximations of this number \cite{Knauf_2015} using Monte-Carlo simulations, both show that there are exponentially many triangulations in terms of the system size $M\times N$.
So lattice triangulations are extensive and can be treated as a well-defined statistical system.
Additionally, the convergence of Glauber dynamics on lattice triangulations was examined in \cite{Caputo_2015,Stauffer_2015} for different parameter sets.

A graph can be constructed from a triangulation by using the 0-simplices (vertices) as graph nodes and the 1-simplices (edges) as graph edges.
In the notation of graph theory triangulations are planar, which is trivial since the graphs are defined in the actual embedding into an euclidean plane, they are even maximal planar with respect to the convex hull of the vertices, i.e. that by including an arbitrary other edge between two nodes that does not leave the convex hull of the vertices the planarity of the triangulation graph is violated.

In order to calculate ensemble averages using Markov-chain Monte Carlo simulations some elementary moves have to be defined that are able to sample all triangulations of a given point set ergodically, so that every triangulation can  be reached from every other triangulation in terms of these moves.
For triangulations Pachner flips \cite{Pachner_1986} can be used as such an elementary move.
For a Pachner flip select an interior edge $\{x_i, x_j\}$ of the triangulation.
This edge is a common edge of the two neighboring triangles $\{x_i, x_j, x_k\}$ and $\{x_i, x_j, x_l\}$.
If the quadrangle $\{x_i, x_j, x_k, x_l\}$ is convex (in lattice triangulations, this is equivalent with the quadrangle being a parallelogram), the Pachner flip is the replacement
\begin{equation}\label{eq:pachner_flip}
  \{x_i, x_j, x_k\}, \{x_i, x_j, x_l\} \rightarrow \{x_i, x_k, x_l\}, \{x_j, x_k, x_l\}.
\end{equation}
If the quadrangle is not convex (i.e. there is a flat or concave angle), the flip is not executable and must be discarded.
Such an exclusion of a flip due to non-convexity does not occur in topological triangulations as considered in \cite{Kownacki_2004,Aste_2012}, because there only topological degrees of freedom are considered, i.e. the connection structure of the vertices; the actual coordinates of the vertices and therewith concavity and convexity depend on the actual embedding and cannot be considered.
(Note that in topological triangulations flips that would include already present edges must be excluded.)
In Fig.~\ref{fig:pachner_move} some executable and non-executable flips are displayed in a small lattice triangulation.
The ergodicity of the Pachner flips \eqref{eq:pachner_flip} for triangulations of arbitrary two-dimensional point sets was proven in \cite{Lawson_1972}, they also conserve the unimodularity of the triangulations as well as the number of vertices and edges.
There are additional Pachner moves \cite{Pachner_1986} that consist of inserting or removing vertices of a triangulation, but these are not needed for ergodicity in our setup and lead to non-full triangulations which are no longer unimodular.

To be able to use ensembles of triangulations that are weighted in terms of their order or disorder, we introduce the energy function
\begin{equation}\label{eq:energy}
  E(\mathcal T) := J\sum_{v \in \mathbf A} \left( k_v^{(\mathcal T)} - k_v^{(\mathcal T_0)} \right)^2
\end{equation}
for a triangulation $\mathcal T$ with $k_v^{(\mathcal T)}$ being the number of incident edges at vertex $v$.
The triangulation $\mathcal T_0$ is a reference triangulation and at the same time the ground state of the energy function.
Throughout the paper we use the maximal ordered triangulation as reference triangulation, which is displayed in Fig.~\ref{fig:triang_examples}.
The introduction of the reference term $k_v^{(\mathcal T_0)}$ may seem somewhat randomly, but if one uses an energy function $\propto \sum_{v \in \mathbf A}  k_v^{(\mathcal T)}$ the contribution of boundary vertices would make a disordered triangulation the ground state of the energy function.
This would conflict with the energy measuring order and disorder of a system.
Another interpretation of introducing the maximal ordered triangulation as reference state is that it introduces a fixed layer of triangles around the miscellaneous triangulations (see Fig.~\ref{fig:triang_examples}) that cannot be flipped and whose vertices are not taken into account for calculating the energy, which leads to fixed boundary conditions.
Note that it is not possible to use periodic boundary conditions (i.e.\,using a triangulation on a torus) for embedded triangulations, because the line segment between two points is not defined uniquely on a torus, and so one cannot decide whether a quadrangle is convex.
Examples for triangulations with different energy can be found in Fig.~\ref{fig:triang_examples}.

Since all full triangulations of a given point set have the same number of vertices, edges and triangles, and since the number of triangles incident with a vertex depends only on the number of edges incident, the energy function \eqref{eq:energy} is the simplest possible local and polynomial energy function in terms of these topological numbers (taking the linear term would lead to a constant energy function).
The same energy function was used for triangulations in \cite{Krueger_2015,Knauf_2015}, a modification using the average number of adjacent edges instead of the reference term was used in \cite{Aste_1999,Kownacki_2004,Aste_2012} for topological triangulations.
It can also be applied to usual graphs as in \cite{Farkas_2004} and is related to the integrated squared curvature used in the discrete form of the Einstein-Hilbert action in the approaches of (causal) dynamical triangulations \cite{Ambjorn_2005}.

\subsection{Spectral graph theory}\label{subsec:graph_theory}

An undirected simple graph $\mathcal G := (\mathbf A, E)$ is a pair consisting of a set $\mathbf A$ (called vertices) with $n = |\mathbf A|$ elements and a set $E$ (called edges) of two-element subsets of $\mathbf A$. 
A triangulation can be interpreted as a graph using the point set $\mathbf A$ as vertices and the 1-simplices of the triangulation as edges $E$.

Important properties of graphs can be found by examining spectral properties (the set of eigenvalues) of matrices associated with graphs.
In the literature mostly the following three $n \times n$-matrices (with $n$ being the number of vertices) are considered for a graph $\mathcal G$ \cite{Chung_2006}:
\begin{itemize}
  \item The \emph{adjacency matrix} $A(\mathcal G)$ with
    \begin{equation*}
      A(\mathcal G)_{ij} := \begin{cases} 1 & \{v_i, v_j\} \in E \\ 0 & \{v_i, v_j\} \notin E \end{cases}
    \end{equation*}
    is a traceless, symmetric matrix that indicates whether two different vertices are connected.
    The matrix elements $A(\mathcal G)_{ij}^k$ equal the number of paths from $v_i$ to $v_j$ containing exactly $k$ edges.
    We will denote its sorted eigenvalues by $\alpha_0 \leq \alpha_1 \leq \dots \leq \alpha_{n - 1}$.
  \item The \emph{Laplacian matrix} $L(\mathcal G) := D(\mathcal G) - A(\mathcal G)$ with degree matrix
    \begin{equation*}
      D(\mathcal G)_{ij} := \delta_{ij} k_{v_i}
    \end{equation*}
    is a discretization of the usual Laplace operator $\vec \nabla^2$.
    The Laplacian matrix is symmetric and positive-semidefinite, the smallest eigenvalue is always 0.
    We denote the sorted eigenvalues of the Laplacian matrix by $\lambda_0 \leq \lambda_1 \leq \dots \leq \lambda_{n - 1}$.
    The multiplicity of the eigenvalue $0$ is the number of connected components of the graph.
  \item The \emph{normalized Laplacian matrix} 
    \begin{equation*}
      \mathcal L(\mathcal G) := \mathbbm 1 - D(\mathcal G)^{-1/2} A(\mathcal G) D(\mathcal G)^{-1/2}
    \end{equation*}
    is useful for describing random walks on arbitrary geometries. 
    This matrix is not considered in this paper, but our calculations can be extended simply to the normalized Laplacian.
\end{itemize}

The spectrum $\mathrm{spec}_M(x)$ of a $n \times n$-matrix $M$ with eigenvalues $\mu_0, \dots, \mu_{n - 1}$ is given by the distribution
\begin{equation*}
  \mathrm{spec}_M(x) := \frac{1}{n} \sum_{i = 0}^{n-1} \delta(x - \mu_i)
\end{equation*}
For large graphs ($n \rightarrow \infty$) the spectrum can be approximated by a continuous function.

A well known result is that the spectrum of the Laplacian matrix can be used to calculate the number of spanning trees, which is $\lambda_1 \cdot \lambda_2 \cdot \dots \cdot \lambda_{n-1} / n$ \cite{Kirchhoff_1847}.
The second-smallest eigenvalues $\lambda_1$ of the Laplacian spectrum is called \emph{algebraic connectivity}, the largest eigenvalue $\lambda_{n-1}$ is called \emph{spectral radius}.
The algebraic connectivity is 0 if and only if the graph is not connected, it can be shown that it increases if one inserts additional vertices into a graph, so it is in fact a good measure for the connectedness of a graph.
Additionally, it can be used to bound other quantities of graphs that are more difficult to calculate, e.g. the isoperimetric number \cite{Mohar_1989}.
Fiedler \cite{Fiedler_1973} and Mohar \cite{Mohar_1991} state that there are the following lower and upper bounds for the algebraic connectivity $\lambda_1$ of a graph $\mathcal G$:
\begin{equation}\label{eq:bounds_algebraic_connectivity}
  2 \eta(\mathcal G) \left[ 1 - \cos\left(\frac{\pi}{n} \right) \right] < \lambda_1(\mathcal G) < v(\mathcal G)
\end{equation}
Here $\eta(\mathcal G)$ is the edge connectivity and $v(\mathcal G)$ is the vertex connectivity, the smallest number of edges or vertices (with incident edges) that must be removed from the graph to create at least two non-connected components.
For the spectral radius $\lambda_{n-1}$ there are the following general bounds (\cite{Mohar_1991,Yu_2004}) for a graph $\mathcal G$:
\begin{equation}\label{eq:bounds_spectral_radius}
  \frac{n}{n - 1} \max\left\{ k_i \mid v_i \in \mathbf A \right\} \leq \lambda_{n - 1}(\mathcal G) \leq \max\left\{ k_i + k_j \mid \{v_i, v_j \} \in \mathcal G \right\}
\end{equation}
where $k_i$ is the degree (number of incident edges) of vertex $v_i$.

One can show that the algebraic connectivity is proportional to the inverse of the synchronization time in consensus dynamics on networks \cite{Almendral_2007}. 
Additionally, for the return probability it governs the large-time in classical and the small frequency behavior in quantum random walks on networks, whereas the spectral radius governs short-time, respectivly, the large frequency behavior \cite{Muelken_2011}. It is also possible to find a bound for the sum of the $j$ largest eigenvalues in terms of the $k$-largest vertex degrees \cite{Grone_1995}.
The sum of the exponentiated eigenvalues of the adjacency and the Laplacian matrix is known as the (Laplacian) Estrada index and has many applications in the study of chemical molecules \cite{Estrada_2000,Estrada_2005,Du_2011}.

The eigenvectors of the Laplacian matrix can be used for examining localization on graphs in terms of the inverse participation ration (IPR) $\chi_m$ of the normalized eigenvector $\ket{m}$ corresponding to eigenvalue $\lambda_{m}$ \cite{Biroli_1999,Monasson_1999}, which is given by
\begin{equation}\label{eq:ipr}
  \chi_m := \sum_k \left( \ket{m}_k \right)^4,
\end{equation}
where $\ket{m}_k$ denotes the $k$-th component of the eigenvector.
Since the normal basis corresponds to the vertices of the graph, these components are the overlap of the eigenvectors and the vertices, so large IPRs correspond to strong localization, whereas small IPRs correspong to delocalization of the corresponding eigenvector.
To examine the localization properties for lattice triangulations, we consider the IPR $\chi_1$ of the algebraic connectivity, the IPR $\chi_{MN-1}$ of the spectral radius and the average IPR $\overline \chi := (\sum_m \chi_m)/MN$.

\subsection{Random graphs}\label{subsec:random_graphs}

In this paper we compare the graph interpretation of unimodular lattice triangulations with some common random graph models.
Similar to \cite{Krueger_2015}, where graph measures as the clustering coefficient (which is the average ratio of the numbers of actual present and possible edges between the neighbors of a node) and the average shortest path length (the shortest path length between two nodes is the minimal number of edges that have to be run through to go from the one to the other node within the graph) were compared to these random graph models, we choose the respective parameters of the random graph models such that the average numbers of nodes and edges are equal to the number $MN$ of triangulation vertices and to the number $3MN - 2(M+N) + 1$ of triangulation edges for comparing the spectral properties.
In the following we describe shortly the different models of random graphs we are comparing the lattice triangulation graphs to, as well as our choice of parameters for these random graphs.

\paragraph{\erdoes random graph} The earliest model of a random graph is the \erdoes random graph $\mathcal G_{n,p}$ with $n$ vertices and the probability $p$ for two distinct vertices to be connected \cite{Erdoes_1959,Erdoes_1960,Erdoes_1961}.
Since the average number of edges in this random graph is $pn(n-1)/2$ we choose the parameters
\begin{equation}\label{eq:erdoes_renyi_prob}
  n = MN \text{ and } p_{ER} = \frac{3MN - 2(M + N) + 1}{MN(MN - 1)/2} \xlongrightarrow{M,N\rightarrow\infty} \frac{6}{MN}
\end{equation}
to get an \erdoes random graph with a comparable number of nodes and edges.

Erd\"os and R\'{e}nyi \cite{Erdoes_1959} categorised the random graphs $\mathcal G_{n,p}$ according to the asymptotics of the average number $np$ of edges incident with a vertex.
For our choice of parameters \eqref{eq:erdoes_renyi_prob} with $np \rightarrow 6$ they found that the graph has one giant component with $n\cdot(1 - 2.5\cdot 10^{-3})$ vertices on average, and $n\cdot 0.25\cdot 10^{-3}$ connected components for $np=6$ and $n\rightarrow \infty$.
The latter result implies that the average \erdoes random graph is not connected for a sufficiently high number of vertices for our choice of parameters, in contrast to the considered lattice triangulations which are always connected.

\paragraph{Watts-Strogatz and Newman-Watts random graph}
Another common random graph model was proposed by Newmann and Watts \cite{Newman_1999}.
It consists of a regular graph on a periodic lattice of $n$ vertices (a ring in one dimension), where each vertex is connected with its $L = 2 k \cdot d$ nearest neighbors (with $d$ being the dimension of the lattice and integer $k$), superimposed with an \erdoes random graph with connection probability $p$.
Numerical calculations for a similar model \cite{Watts_1998} showed that there is a crossover region in the parameter $p$ where this model shows a small-world (but not scale-free) behavior.

For a Newman-Watts random graph with $n = MN$ nodes the requirement of the average number of edges to match the number of triangulation edges implies the rewiring probability $p$:
\begin{equation*}
  |E| = \frac{1}{2}LMN + \frac{1}{2}pMN(MN - L - 1) \Rightarrow p = \frac{(6-L)MN - 4(M+N) + 2}{MN(MN-L-1)}
\end{equation*}
Since no edges are deleted in the Newman-Watts model, the number of edges in the original regular lattice must be smaller than the number of edges in the triangulation.
For the 1d Newman-Watts model this implies $L = 2$ or $L = 4$, for the 2d model one has to choose $L = 4$.
In the further considerations we will only use the 1d case with $L = 4$, because for the two other cases the original regular lattice has only edges to the nearest neighbors and a clustering coefficient $C(p=0) = 0$, which differs from the behavior of triangulations.

\paragraph{\barabasi random graph}
Barab\'{a}si and Albert \cite{Barabasi_1999} proposed a preferential attachment random graph model with the following construction in order to achieve a power-law degree distribution:
Start from a graph of $m$ vertices without edges.
Then execute $t$ of the following steps:
Insert one vertex and edges from this vertex to $m$ existing vertices so that the probability $p_i$ for an edge between vertex $v_i$ and the new vertex is proportional to the node degree $k_i$.
Hence vertices with high vertex degree have a high probability to be linked to the new vertex (preferential attachment).
The resulting graph consists of $n = m + t$ vertices and $m\cdot t$ edges.

To compare the \barabasi random graph to a random triangulation on an $M\times N$ integer lattice, we choose $m = 3$ and $t = MN - 3$ so that the corresponding \barabasi random graph has $3MN$ vertices and $MN - 9$ edges.
The number of vertices matches the number of triangulation vertices, for $M,N \gg 1$ also the number of edges match.

%% file: analytical_solution.tex
\section{Analytical solution}\label{sec:analytical_solution}

If one extends the ground state to periodic boundaries (resulting in a periodic, triangular lattice), also analytical solutions for the eigenvalues of the adjacency and the Laplacian matrix can be found.
It is even possible to use perturbation theory to calculate the effect of flips on the spectrum and therewith to approximate the spectral properties of low-energy triangulations.

\paragraph{Adjacency and Laplacian spectrum (ground state)}

\begin{figure}
  \includegraphics{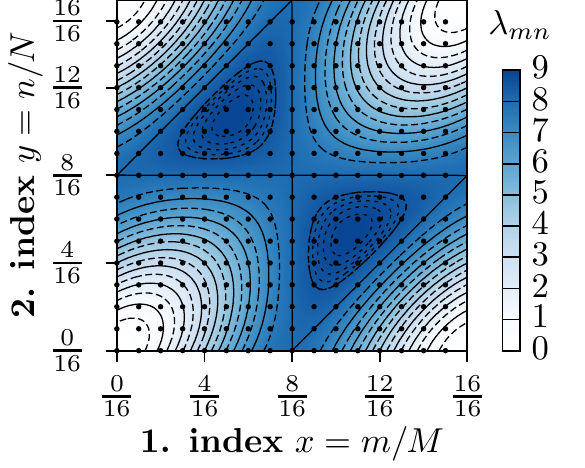}
  \includegraphics{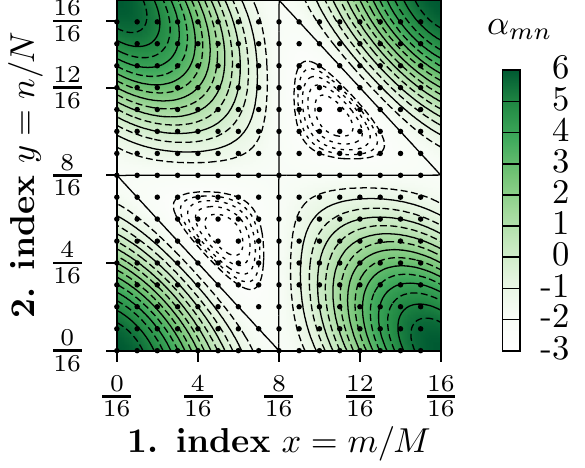}
  \caption{\label{fig:analytical_spectrum}(Color online) Analytical calculated eigenvalues of the Laplacian matrix (left, Eq. \eqref{eq:periodic_eigenvalues}) and of the adjacency matrix (right, Eq. \eqref{eq:periodic_eigenvalues_adjacency}) for the periodic ground state of a triangulation in terms of $m/M$ and $n/N$. The solid lines are the isocurves for the integer values, the dashed lines are the isocurves for the half-integers and the dots represent the realised values for $M=N=16$. For larger triangulations the dots become denser.}
\end{figure}

In this section we calculate the spectrum of the adjacency and the Laplacian matrix for a triangular $M\times N$-lattice graph with periodic boundary conditions.
This is an approximation for the ground state of a triangulation.
The components $A_{(i_1 j_1)(i_2 j_2)} := A_{i_1\cdot N + j_1, i_2\cdot N + j_2}$ (with $i_k \in \mathbbm Z_M$, $j_k \in \mathbbm Z_N$) of the adjacency matrix are given by
\begin{equation}\label{eq:adjacency_periodic_gs}
  \begin{split}
    A_{(i_1 j_1)(i_2 j_2)} &= \delta^{(M)}_{i_1 i_2} B_{j_1 j_2} + \delta^{(M)}_{i_1 (i_2 + 1)} C_{j_1 j_2} + \delta^{(M)}_{(i_1 + 1) i_2} C_{j_2 j_1} \\
    B_{j_1 j_2} &= \delta^{(N)}_{j_1 (j_2 + 1)} + \delta^{(N)}_{(j_1 + 1) j_2} \\
    C_{j_1 j_2} &= \delta^{(N)}_{j_1 j_2} + \delta^{(N)}_{(j_1 + 1) j_2}
  \end{split}
\end{equation}
The matrix $B$ describes the connections between vertices in a common row, the matrix $C$ describes the connections between the different rows.
$\delta^{(M)}_{i_1 i_2}$ is the $M$-periodic Kronecker delta (1 for $i_1 = i_2 + k\cdot M$ with $k \in \mathbb Z$, 0 otherwise).
The Laplacian matrix $L$ is then given by
\begin{equation*}
  L_{(i_1 j_1)(i_2 j_2)} := 6\delta^{(N)}_{j_1 j_2}\delta^{(M)}_{i_1 i_2} - A_{(i_1 j_1)(i_2 j_2)}
\end{equation*}

Using a decomposition into Fourier components one can show that $\alpha_{mn} \in \mathbb R$ and $\ket{m,n} \in \mathbb C^{MN}$ solve the eigenvalue equation
\begin{equation*}
  A \ket{m,n} = \alpha_{mn} \ket{m,n}
\end{equation*}
with eigenvalues
\begin{equation}\label{eq:periodic_eigenvalues_adjacency}
  \begin{split}
    \alpha_{m,n} = 2\left[ \cos\left( 2\pi \frac{n}{N}\right) + \cos\left( 2 \pi \frac{m}{M}\right) + \cos\left( 2\pi \left(\frac{m}{M} + \frac{n}{N} \right)\right)\right]
  \end{split}
\end{equation}
and the corresponding components of the eigenvectors
\begin{equation}\label{eq:periodic_eigenvectors}
  \ket{m,n}_{kl} = \frac{1}{\sqrt{MN}} \exp\left( 2\pi i \frac{k\cdot m}{M} \right) \exp\left( 2\pi i \frac{l\cdot n}{N} \right).
\end{equation}
The eigenvalues and (up to a scaling factor) the eigenvectors depend only on the relative indices $m/M$ and $n/N$. 
So the spectrum as visualized in Fig.~\ref{fig:analytical_spectrum} is independent of the system size in terms of the relative indices, but the actual system size becomes important to determine which relative indices $m/M$ and $n/N$ with $m\in \mathbb Z_M, n \in \mathbb Z_N$ are possible.
The two-fold degenerated smallest eigenvalues of the adjacency matrix are
\begin{equation*}
  \alpha_0 = \alpha_1 = \alpha_{\frac{M}{3},\frac{N}{3}} = \alpha_{\frac{M}{3},\frac{N}{3}} = -3
\end{equation*}
the largest eigenvalue is $\alpha_{MN-1} = \alpha_{0,0} = 6$.

The eigenvectors of the adjacency matrix \eqref{eq:periodic_eigenvalues_adjacency} also solve the eigenvalue equation for the Laplacian matrix
\begin{equation*}
  L \ket{m,n} = \lambda_{m,n} \ket{m,n}
\end{equation*}
with eigenvalues
\begin{equation}\label{eq:periodic_eigenvalues}
  \begin{split}
    \lambda_{m,n} &= 6 - 2\left[ \cos\left( 2\pi \frac{n}{N}\right) + \cos\left( 2 \pi \frac{m}{M}\right) + \right. \\
    &+\left. \cos\left( 2\pi \left(\frac{m}{M} - \frac{n}{N} \right)\right)\right]
  \end{split}
\end{equation}
which are visualized in Fig.~\ref{fig:analytical_spectrum}.
These Laplacian eigenvalues of a periodic triangular lattice were also found in \cite{Heuvel_2001}.
The lowest eigenvector $\lambda_0 = \lambda_{0,0}$ is zero since the triangulation is connected. The algebraic connectivity $\lambda_1$ is
\begin{equation}\label{eq:periodic_gs_algebraic_connectivity}
  \lambda_1 = 
  \begin{cases}
    \lambda_{1,0} = \lambda_{N-1,0} = 6 - 6 \cos \left( \frac{2\pi}{N} \right) & N\geq M \\
    \lambda_{0,1} = \lambda_{0,M-1} = 6 - 6 \cos \left( \frac{2\pi}{M} \right) & N < M
  \end{cases}
\end{equation}
and for $M, N \in 3 \cdot \mathbb{N}$ the spectral radius $\lambda_{MN-1}$ is
\begin{equation}\label{eq:periodic_gs_spectral_radius}
  \lambda_{MN - 1} = \lambda_{\frac{N}{3},\frac{2N}{3}} = \lambda_{\frac{2N}{3},\frac{N}{3}} = 9
\end{equation}

The absolute value of all components of the eigenvectors \eqref{eq:periodic_eigenvectors} is $(MN)^{-1}$, this implies that the inverse participation ratio of every eigenvector and also the average IPR takes its minimal value $\overline \chi = \chi_m = (MN)^{-1}$.

\paragraph{Adjacency and Laplacian spectrum (low energy states)}
As a next step we use the usual quantum mechanical perturbation theory to calculate analytically the influence of a single and multiple flips on the eigenvalues of the Laplacian operator.
Consider a flip of the edge $\{(a,b)(a-1,b+1)\}$ into the new edge $\{(a-1,b)(a,b+1)\}$.
The Laplacian operator $L^\prime = L + V$ of the flipped state can be calculated using the Laplacian operator $L$ of the ground state and a perturbation matrix $V$ with the following non-vanishing components:
\begin{equation*}
  \begin{split}
    V_{ab,ab} &= V_{(a-1)(b+1),(a-1)(b+1)} = -1 \\
    V_{(a-1)b,(a-1)b} &= V_{a(b+1),a(b+1)} = 1 \\
    V_{ab,(a-1)(b+1)} &= V_{(a-1)(b+1),ab} = 1 \\
    V_{(a-1)b,a(b+1)} &= V_{a(b+1),(a-1)b} = -1
  \end{split}
\end{equation*}
The expectation value for the perturbation $V$ is then:
\begin{equation*}
  \begin{split}
    & \braket{m,n\mid \hat V \mid m',n'} = \frac{1}{MN} e^{2 \pi i \frac{(m' - m) a}{M}} e^{2 \pi i \frac{(n' - n) b}{N}} \\
    & \cdot\left[ 4e^{\pi i \frac{m - m'}{M}} e^{\pi i \frac{n' - n}{N}}\sin\left(\pi \frac{m-m'}{M}\right)\sin\left(\pi \frac{n'-n'}{N}\right) + \right. \\
    & + \left. \left( e^{2\pi i \frac{n'}{N}} - e^{-2 \pi i \frac{n}{N}}\right) \left( e^{-2\pi i \frac{m'}{M}} - e^{2 \pi i \frac{m}{M}}\right) \right]
  \end{split}
\end{equation*}

In the following we consider only quadratic lattices $M=N$, the rectangular case can be treated in an analogue fashion. 
For calculating the first order correction for the spectral radius one has to take into account that $\lambda_{N/3,2N/3} = \lambda_{2N/3,N/3}$ is two fold degenerate, so one has to calculate the perturbation matrix and to diagonalize it.
The relevant terms of the perturbation matrix are the diagonal element
\begin{equation*}
  \Braket{\frac{N}{3},\frac{2N}{3} | \hat V | \frac{N}{3}, \frac{2N}{3}} = \frac{3}{N^2}
\end{equation*}
and the complex off-diagonal element
\begin{equation*}
  \Braket{\frac{N}{3},\frac{2N}{3} | \hat V | \frac{2N}{3}, \frac{N}{3}} = \frac{3}{N^2}e^{\frac{2 \pi i}{3}(a - b - 1)}
\end{equation*}
The eigenvalues of the resulting hermitian $2\times 2$-matrix are $0$ and $6 / N^2$, so in first order perturbation theory the spectral radius becomes
\begin{equation}\label{eq:perturbation_theory_spectral_radius}
  \lambda_{N^2 - 1} \approx 9 +  \frac{6}{N^2}
\end{equation}
with the second being the important contribution since the spectral radius needs to be the biggest eigenvalue.
This relation will be used later on to estimate the energy dependence of the spectral radius for small energies.

The perturbation theory unfortunately fails if one considers the algebraic connectivity.
For the case $M = N$ the corresponding eigenvalue is six-fold degenerated and the perturbation matrix has to be diagonalized numerically.
This leads for all $M,N$ to a decrease in the algebraic connectivity, which is consistent with a direct diagonalization of the new Laplacian $L + V$ for the periodic case.
In contrast to that, if one diagonalizes $L + V$ directly for the non-periodic case, the algebraic connectivity increases.
So for the small eigenvalues the difference between a regular and a periodic triangulation leads to a qualitative change in the behavior of the algebraic connectivity, which is quite comprehensible since the algebraic connectivity is dominated by the vertices with low degree (two for non-periodic and six for periodic triangulations), which can be seen already in the bounds \eqref{eq:bounds_algebraic_connectivity}.
The spectral radius is determined by the vertices with high degree, which is six for both periodic and non-periodic triangulations, also apparent in the bounds \eqref{eq:bounds_spectral_radius}.

%% file: results.tex
\section{Triangulation ensembles as random graph models}\label{sec:triang_spectra}
In this section we calculate spectral graph observables (Laplacian and adjacency spectrum, algebraic connectivity and spectral radius) numerically for various ensembles of triangulations.
Other graph observables as the degree distribution, the clustering coefficient and the average minimal path length were calculated before in Ref.\,\cite{Krueger_2015}.
Let $\Omega_{\mathbf{A}}$ be the set of all possible triangulations of the given vertices $\mathbf A$.
An ensemble is defined as a map $P:\Omega_{\mathbf{A}} \rightarrow [0,1]$  that assigns every possible triangulation of $\mathbf A$ a weight so that $\sum_{\mathcal T \in \Omega_{\mathbf{A}}} P(\mathcal T) = 1$. 
An ensemble average of an observable $\mathcal O$ can then be calculated using 
\begin{equation}\label{eq:ensemble_average}
  \langle \mathcal O \rangle := \sum_{\mathcal T \in \Omega_{\mathbf{A}}} P(\mathcal T) \mathcal O(\mathcal T)
\end{equation}
This notion of ensembles was already applied to random graphs in \cite{Farkas_2004}.

In this paper we consider three different ensembles that are well-known in statistical physics:
\begin{itemize}
  \item The \emph{random ensemble} assigns to each triangulation the constant probability $P = 1/|\Omega_{\mathbf A}|$ and can be viewed as a democratic sum since each triangulation contributes in equal measure. 
    This ensemble is independent of the definition of a triangulation energy.
    Ensemble averages of the observable $\mathcal O$ in the random ensemble will be denoted by $\langle \mathcal O \rangle_{\mathrm{rnd}}$ or simply $\langle \mathcal O \rangle$
  \item The \emph{microcanonical ensemble} at constant energy $E$ assigns every triangulation with energy $E$ a constant probability, and every triangulation with a different energy the probability $0$. 
    This ensemble is a random ensemble in the subset of all triangulations $\Omega_{\mathbf A}|_E := \{ \mathcal T \in \Omega_{\mathbf A} \mid E(\mathcal T) = E \}$ with energy $E$.
    Microcanonical ensemble averages of an observable $\mathcal O$ at energy $E$ will be denoted by $\langle \mathcal O \rangle_{\mathrm{mc}}(E)$
  \item The \emph{canonical ensemble} at constant inverse temperature $\beta$ assigns every triangulation $\mathcal T$ with energy $E$ the Boltzmann probability $P(\mathcal T) = \exp(-\beta E) / Z$ with $Z = \sum_{\mathcal T \in \Omega_{\mathbf A}} \exp(-\beta E(\mathcal T))$.
    Formally one can define the canonical ensemble also for negative temperatures $\beta < 0$.
    In the case of triangulations with the considered energy function \eqref{eq:energy} negative temperatures correspond to the usual positive temperatures with negative coupling constant $J < 0$.
    The canonical ensemble at infinite temperature $(\beta = 0)$ equals the ensemble of random triangulations.
    Canonical ensemble averages of an observable $\mathcal O$ at inverses temperature $\beta$ will be denoted by $\langle \mathcal O \rangle_{\mathrm{c}}(\beta)$
\end{itemize}

Usually the set of all triangulations $\Omega_{\mathbf A}$ of the vertices $\mathbf A$ is too big to generate, so calculating the ensemble average using Eq.~\eqref{eq:ensemble_average} is impossible.
So one has to restrict to a suitable subset of all triangulations to approximate the ensemble average.
For the random ensemble one can simply generate randomly a certain number of triangulations as the desired subset (simple sampling), for other ensembles the problem of the simple sampling procedure is that in most cases only triangulations with low weights $P(\mathcal T)$ are encountered, and the probability to find a highly contributing triangulation is low.
One better uses an importance sampling algorithm that preferentially selects triangulations with a high ensemble weight.

The most common possibility called Metropolis Monte Carlo \cite{Metropolis_1953} is to create a sequence of triangulations, a so-called Markov chain, so that the probability for a triangulation to appear in the Markov chain equals the weight of this triangulation in the ensemble.
For the triangulation one generates the sequence by executing single Pachner moves that transform a triangulation $\mathcal T_1$ into a triangulation $\mathcal T_2$ with a probability $P(\mathcal T_1 \rightarrow \mathcal T_2)$ that is chosen so that
\begin{equation*}
  \frac{P(\mathcal T_1 \rightarrow \mathcal T_2)}{P(\mathcal T_2 \rightarrow \mathcal T_1)} = \frac{P(\mathcal T_1)}{P(\mathcal T_2)}
\end{equation*}
This equation commonly is denoted as the detailed balance condition.
The step probability 
\begin{equation*}
P(\mathcal T_1 \rightarrow \mathcal T_2) = S(\mathcal T_1 \rightarrow \mathcal T_2) \cdot A(\mathcal T_1 \rightarrow \mathcal T_2)
\end{equation*}
is usually split into the probability $S(\mathcal T_1 \rightarrow \mathcal T_2)$ for selecting a given step $\mathcal T_1 \rightarrow \mathcal T_2$ and the probability $A(\mathcal T_1 \rightarrow \mathcal T_2)$ for accepting a previously selected step.
If choosing the Pachner moves by choosing the flip edge uniformly from the set of all interior edges and using $A = 0$ for non-executable flips, the contribution of the selection probabilities cancels and one can use the standard Metropolis choice for the acceptance probabilities \cite{Metropolis_1953}:
\begin{equation}\label{eq:acceptance_probability_metropolis}
  A(\mathcal T_1 \rightarrow \mathcal T_2) = \min \left[1, \frac{P(\mathcal T_2)}{P(\mathcal T_1)} \right]
\end{equation}
An important requirement for the use of the Metropolis or similar algorithms is that the steps constructing the Markov chain must be ergodic, i.e. every two possible states of the system must be connected by a finite number of steps with non-vanishing step probability.
The requirement is fulfilled for Pachner moves in two-dimensional triangulations \cite{Lawson_1972} as long as no step probability is chosen to be zero.

For the random ensemble the Metropolis acceptance probability \eqref{eq:acceptance_probability_metropolis} is always equal to 1 due to the fact that every triangulation has the same weight.
For the canonical ensemble \eqref{eq:acceptance_probability_metropolis} the acceptance probability has the following well-known form:
\begin{equation}
  A(\mathcal T_1 \rightarrow \mathcal T_2) = \min \left[1, e^{- \beta \Delta E(\mathcal T_1 \rightarrow \mathcal T_2)} \right];\quad \Delta E(\mathcal T_1 \rightarrow \mathcal T_2) := E(\mathcal T_2) - E(\mathcal T_1)
\end{equation}
Note that the sampling algorithm has to be modified for the microcanonical ensemble, as described in Sec.\,\ref{subsec:results_microcanonical}.
For the canonical ensemble we also do not use Metropolis, but Wang-Landau sampling, which will be explained in details in Sec.\,\ref{subsec:results_canonical}.

%% file: results_random.tex
\subsection{Random triangulation}\label{subsec:results_random}

In this section we consider spectral graph observables of random lattice triangulations and their scaling behavior in terms of the size of the integer lattice.
The results of these calculations are especially interesting because they do not depend on the choice of the energy function, since every possible triangulation contributes with constant weight.

The spectral observables of the random ensemble are calculated using the Metropolis algorithm \cite{Metropolis_1953} averaging over more than $1000$ different random triangulations.
Due to the fact that every triangulation occurs in the random ensemble with constant probability, the acceptance probability \eqref{eq:acceptance_probability_metropolis} used in the Metropolis algorithm is always 1, so every possible and suggested Pachner move will be executed.
To obtain a stationary distribution of random triangulations, we start with the maximal ordered triangulation and propose $10^5 M\cdot N$ Pachner moves, which is well above the diameter $o(MN(M+N))$ of the flip graph \cite{Caputo_2015} for the considered system sizes.
In order to avoid an influence of the autocorrelation of successive triangulations in the Markov chain, we ensure that the number of steps between two measurements is larger than 100 autocorrelation times, which we measured to be smaller than $10 M\cdot N$ for the considered spectral observables and system sizes, and which was measured for other graph observables before in Ref.\,\cite{Krueger_2015}.

For the calculations in this paper we use only quadratic lattice sizes $N\times N$; considering rectangular lattices is possible but does not change the results much, for large lattices the only relevant parameter is the number of vertices $M\cdot N$ \cite{Krueger_2015}.
The accessible system sizes are rather limited compared to topological triangulations \cite{Kownacki_2004,Aste_2012} for two reasons: 
First, one has to check whether a proposed flip is executable (compare Sec.\,\ref{subsec:lattice_triangulations}), which can be done for two-dimensional lattice triangulations by checking whether the two diagonals have the same mids, or by testing for convexity in more general cases.
These checks consume more computation time for a flip than in the topological case, where only incidences have to be looked up.
Second, every time encountering a non-executable flip the triangulation remains unchanged, which increases e.g. the autocorrelation time of the algorithm.
Note that proposing only executable flips is not suitable, because this makes it difficult to fulfill the detailed balance.
We were able to calculate the spectral properties of random triangulations for system sizes between $4 \times 4$ and $64 \times 64$.

We compare the ensemble averages for random triangulations as graphs with the \erdoesnospc, the Newman-Watts and the \barabasi random graphs with parameters chosen according to Sec.~\ref{subsec:random_graphs}, the generation of these random graphs was done using the NetworkX framework \cite{Hagberg_2008}.

\paragraph{Spectrum of the adjacency matrix}
For the \erdoes random graph there are some analytical results for the spectrum of the adjacency matrix.
The random graph adjacency spectra are strongly related with the spectra of random matrices, which follow in most cases a semi-circle distribution \cite{Wigner_1955,Wigner_1958,Juhasz_1981}.

Numerical investigations showed that for constant edge probability $p \neq 0, 1$ the adjacency spectrum of the \erdoes model converges to the semi-circle distribution up to the largest eigenvalue, but there are additional peaks for $p \propto n^{-1}$, which we actually choose to compare with random triangulations (see \cite{Evangelou_1992,Evangelou_1992b} for investigations on the level of random matrices and \cite{Bauer_2001,Farkas_2001} for investigations on random graphs).

\begin{figure}
  \includegraphics{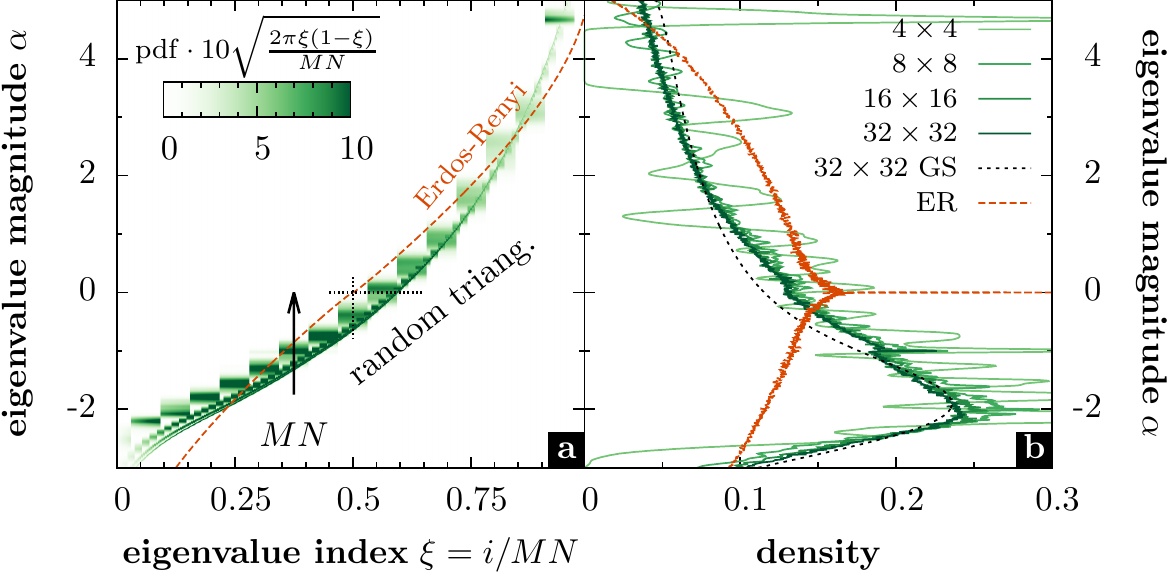}
  \caption{\label{fig:adjacency_spectrum_random_heatmap}(Color online) Average adjacency spectrum of random triangulations for different system sizes. The color scale (a) shows the probability density function (PDF) for every normalized index of the eigenvalues scaled with the maximum value expected from order statistics \eqref{eq:scaling_spectrum_pdf}. The orange, dashed line is the spectrum of the \erdoes random graph with $32^2$ vertices. One sees a convergence of the spectrum for increasing system size. The index-summed PDF (b) is compared with the \erdoes random graph and the spectrum of the maximal ordered triangulation (dashed black line). The dotted lines in the index-resolved spectrum show the cuts that are displayed in Fig.~\ref{fig:adjacency_spectrum_random_cuts}}
\end{figure}

\begin{figure}
  \includegraphics{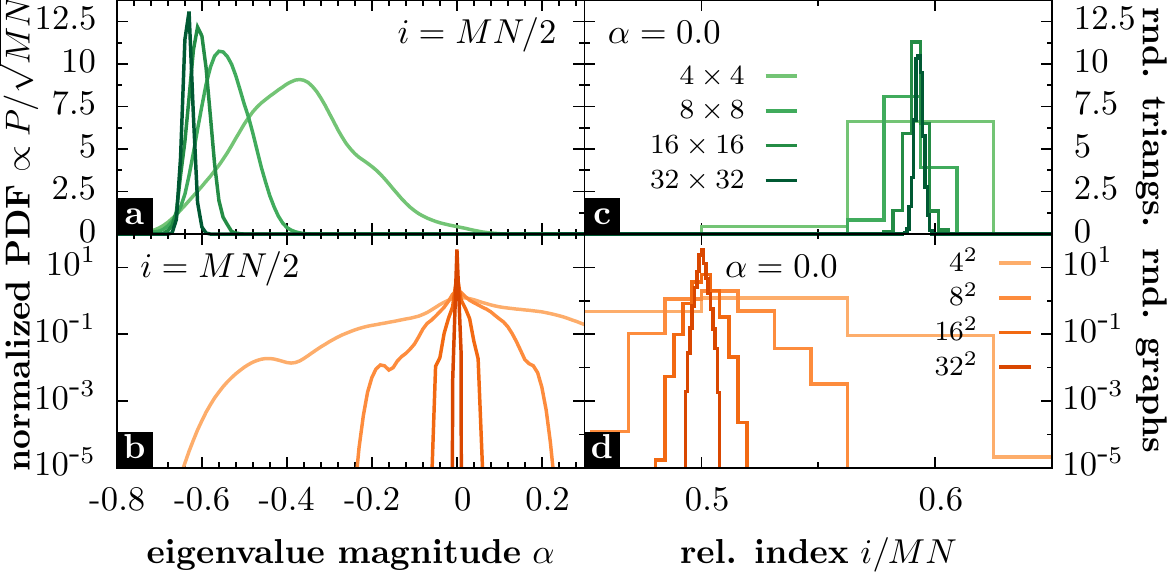}
  \caption{\label{fig:adjacency_spectrum_random_cuts}(Color online) Cuts through the probability density function (PDF) of the adjacency spectrum of random triangulations (a,c) and \erdoes random graphs (b,d). The cut through index $i = MN/2$ (a,b) displays the probability distribution of the eigenvalues for this index, the cut through $\alpha = 0$ (c,d) shows the probability for an index to have this eigenvalue magnitude.}
\end{figure}

In Fig.~\ref{fig:adjacency_spectrum_random_heatmap} the probability density function (PDF) for the spectrum of the adjacency matrix of random triangulations are displayed in an index-resolved and an index-summed way for different lattice sizes.
The spectrum is compared with the spectra of the maximal ordered triangulations and the \erdoes random graph.
Fig.~\ref{fig:adjacency_spectrum_random_cuts} displays cuts through the random triangulation and the \erdoes adjacency PDF for the eigenvalue index $i = MN/2$ and the eigenvalue magnitude $\alpha = 0$.
The displayed PDFs are created using a kernel density estimation \cite{Rosenblatt_1956,Parzen_1962} with Gaussian kernel and Silverman's rule \cite{Silverman_1986} for the width of the Gaussian.

The adjacency spectra of random and maximal ordered triangulations both have a main peak around the eigenvalue magnitude of -2. 
The only difference is that at $\alpha \approx 0$ the random PDF has a higher value than the maximal ordered one, and vice versa at $\alpha \approx 4.5$.

In contrast to the random triangulations the numerically calculated adjacency spectrum of the \erdoes random graph is symmetric with respect to $\alpha \approx 0$ and behaves like Wigner's semicircle law with an additional peak at $\alpha = 0$.
The peak is due to the eigenvalue $\alpha_{MN/2} = 0$ occurring in all random graphs that were calculated numerically and can also be seen as a small horizontal piece in the index-resolved adjacency PDF. 

Considering the values of the index-resolved PDF one can see for each eigenvalue index a peaked distribution of the eigenvalue magnitudes with increasing height and decreasing width for growing system size.
One can assume that this is due to the number of eigenvalues $MN$ growing with the lattice sizes, but all having values in a fixed interval $\approx [-4,6]$.
This behavior can be quantified using order statistics:
Assume that one draws $n$ random variables $X_i$ from a uniform distribution on the fixed interval $[a,b]$ and sorts them so that $X_i \leq X_j$ for $i < j$. 
Then the probability density function (PDF) for the $X_k$ is given by the beta distribution
\begin{equation*}
  P(X_k = x) = \frac{1}{b-a} \frac{n!}{k! (n - k - 1)!} x^k (1 - x)^{n - k - 1}
\end{equation*}
The PDF takes its maximal value at $x_{\mathrm{max}} = a + (b-a)k / (n - 1)$, the maximal value is
\begin{equation}\label{eq:scaling_spectrum_pdf}
  P(X_k = x_{\mathrm{max}}) 
  \overset{n \rightarrow \infty}{\rightarrow} \frac{1}{b-a} \sqrt{\frac{n}{2\pi \xi (1 - \xi)}}
\end{equation}
where $\xi = k / (n-1)$ denotes the relative index and Stirling's formula was used for calculating the asymptotics of the factorials.
So one expects that the maximal value of the PDF scales $\propto (MN)^{0.5}$ independently for each relative index $\xi = k / (n-1)$. 
This is in fact the scaling behavior that can be found in the Fig.~\ref{fig:adjacency_spectrum_random_heatmap} and Fig.~\ref{fig:adjacency_spectrum_random_cuts} for the adjacency spectrum.



\begin{figure}
  \includegraphics{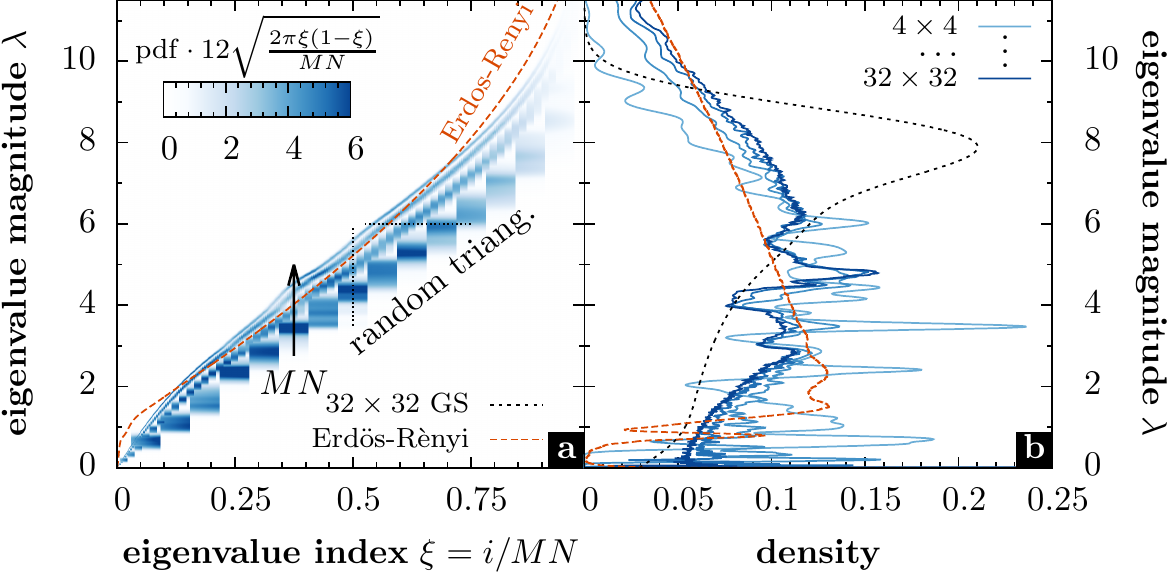}
  \caption{\label{fig:laplacian_spectrum_random_heatmap}(Color online) Average Laplacian spectrum of random triangulations for different system sizes. The color scale (a) shows the probability density function (PDF) for every normalized index of the eigenvalues scaled with the maximum value expected from order statistics \eqref{eq:scaling_spectrum_pdf}. The orange, dashed line is the spectrum of the \erdoes random graph with $32^2$ vertices. One sees a convergence of the spectrum for increasing system size. The index-summed PDF (b) is compared with the \erdoes random graph and the spectrum of the maximal ordered triangulation (dashed black line). The dotted lines in the index-resolved spectrum show the cuts that are displayed in Fig.~\ref{fig:laplacian_spectrum_random_cuts}}
\end{figure}

\begin{figure}
  \includegraphics{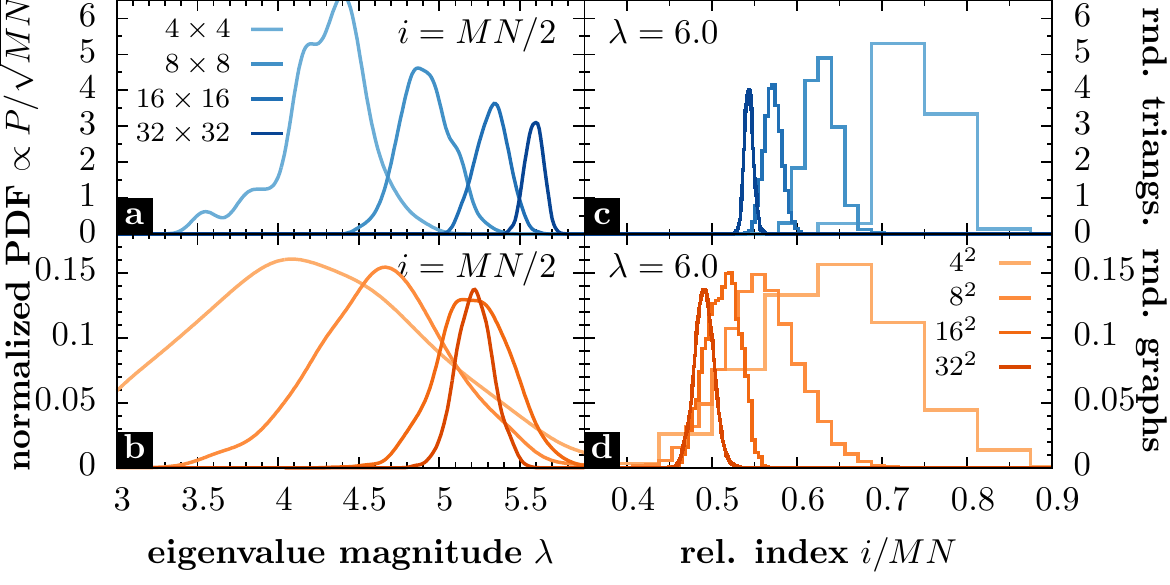}
  \caption{\label{fig:laplacian_spectrum_random_cuts}(Color online) Cuts through the probability density function (PDF) of the Laplacian spectrum of random triangulations (a,c) and \erdoes random graphs (b,d). The cut through index $i = MN/2$ (a,b) displays the probability distribution of the eigenvalues for this index, the cut through $\lambda = 6$ (c,d) shows the probability for an index to have this eigenvalue magnitude.}
\end{figure}

\paragraph{Spectrum of the Laplacian matrix}
The same considerations as for the adjacency spectrum can also be done for the Laplacian spectrum of random triangulations and graphs.
In Fig.~\ref{fig:laplacian_spectrum_random_heatmap} the index-resolved and index-summed probability density function (PDF) for the spectrum of the Laplacian matrix of random triangulations are displayed.
For comparison also the Laplacian spectra of the maximal ordered triangulation and an \erdoes random graph with the same number of vertices are plotted.
Fig.~\ref{fig:laplacian_spectrum_random_cuts} displays cuts through the PDF for a given eigenvalue index and a given eigenvalue magnitude.
As for the spectra of the adjacency matrix, the PDFs are calculated using a Gaussian kernel density estimation and normalized with the order statistics expected maximal values \eqref{eq:scaling_spectrum_pdf}.

If one compares the Laplacian random spectrum with the spectrum of the maximal ordered triangulation two main things differ:
For the maximal ordered triangulation the largest possible eigenvalue is 9 (as shown analytically for the periodic maximal ordered triangulation), whereas the largest eigenvalues of the random triangulations are much higher for increasing lattices sizes, which can be understood in terms of the lower bound \eqref{eq:bounds_spectral_radius} of the spectral radius given by the maximal vertex degree that is higher for random triangulations.
The index-summed PDF is peaked around an eigenvalue magnitude of 8 for the maximal ordered triangulation, for the random triangulation (of lattices with size bigger than $10\times 10$) there is a peak around an eigenvalue magnitude of 5 which is less dominant than the peak of the maximal ordered one.

For the considered parameter set of \erdoes random graphs there are approximative analytical calculations \cite{Dean_2002} for the spectrum of the Laplacian that coincide with earlier numerical calculations \cite{Biroli_1999}.
Ding and Jiang \cite{Ding_2010} showed that the spectrum of the adjacency matrix of a random (Erd\"os-R\'{e}nyi) graph converges to the semi-circle distribution, and the spectrum of the Laplacian matrix converges to a free convolution of the semi-circle distribution and a normal distribution.

The Laplacian spectrum of the Erd\"os-R\'enyi random graph looks similar to the spectrum of the triangulations for eigenvalue magnitudes bigger than 6, both exhibit a linear density decrease for increasing eigenvalue magnitudes.
For eigenvalue magnitudes between 2 and 6 there are several peaks in the spectrum of the random triangulation which also survive if one considers the limit $MN \rightarrow \infty$, whereas the random graph spectrum in this interval is nearly linear.
Between eigenvalue magnitudes 0 and 2 the Laplacian spectrum of the random graph shows two peaks and three dips, while the spectrum of the random triangulation is smooth at most for large lattice sizes.

Comparing the Laplacian spectra cuts of the random triangulations and the \erdoes random graphs with the expectations from the order statistics in Fig.~\ref{fig:laplacian_spectrum_random_cuts}, one can find similar results as in the case of the adjacency spectra.

\paragraph{Algebraic connectivity and spectral radius}
\begin{figure}
  \includegraphics{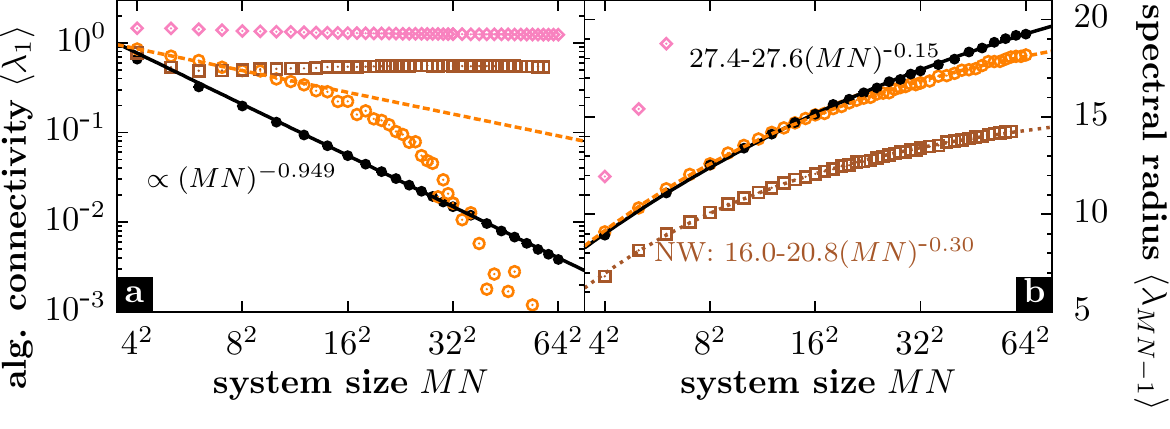}
  \caption{\label{fig:spectrum_observables_random}(Color online) Algebraic connectivity $\lambda_1$ (a) and spectral radius $\lambda_{MN-1}$ (b) of random triangulations (\protect\includegraphics{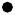}), \erdoes (\protect\includegraphics{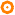}), Newman-Watts (\protect\includegraphics{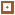}) and \barabasi (\protect\includegraphics{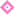}) random graphs for different number of vertices $MN$. The displayed lines are power law fits for the numerical data.}
\end{figure}

In this section we examine the dependence of the smallest Laplacian eigenvalue $\lambda_1$ (algebraic connectivity) and the biggest Laplacian eigenvalue $\lambda_{MN-1}$ (spectral radius) of random triangulations on the lattice size.
The results of the Monte-Carlo simulations for random triangulations can be found in Fig.~\ref{fig:spectrum_observables_random}, as well as the values of the algebraic connectivity and the spectral radius for the different considered models of random graphs.

Using a power law fit for the algebraic connectivity of random triangulations one finds that
\begin{equation*}
  \langle \lambda_1 \rangle \approx (10.7 \pm 0.2) (MN)^{-0.949 \pm 0.003}
\end{equation*}
This compares well with the analytical result for the periodic maximal ordered triangulation \eqref{eq:periodic_gs_algebraic_connectivity}
\begin{equation*}
  \lambda_1^{(\mathrm{p.gs})} = 4 - 4\left[ 1 - \frac{4 \pi^2}{N^2} + \mathcal O\left(\frac{1}{N^4}\right) \right] \propto (M\cdot N)^{-1},
\end{equation*}
because the algebraic connectivity is determined by the edge and vertex connectivity (compare the bounds \eqref{eq:bounds_algebraic_connectivity}), which are themselves determined by the vertices with small degree mainly located at the boundary.
For the algebraic connectivity of the \erdoes random graph one finds a different behavior:
For $MN \lessapprox 200$ there is a power law behavior $\propto x^{-0.44}$, but for $MN \gtrapprox 200$ there is a power law behavior $\approx x^{-2}$.
The change in the power law exponent corresponds to the fact that for large number of vertices $MN$ and $p \rightarrow 6/MN$ some of the random graphs become disconnected as described in Sec.~\ref{subsec:random_graphs} and have an algebraic connectivity $\lambda_1 = 0$.
So the fraction of disconnected random graphs increases with $MN$ which influences the averaging of the algebraic connectivity.
The algebraic connectivities of the Newman-Watts model and the \barabasi model converge to the finite values $\lambda_{1,NW} = 0.553 \pm 0.003$ and $\lambda_{1,BA} = 1.238 \pm 0.001$.

For the spectral radius of the random triangulation the power law fit yields
\begin{equation*}
  \langle \lambda_{MN - 1} \rangle = (27.4 \pm 0.5) - (27.6 \pm 0.3)\cdot (MN)^{-0.147 \pm 0.006},
\end{equation*}
which is much larger than the value $9$ for the maximal ordered triangulation, since the bounds \eqref{eq:bounds_spectral_radius} of the spectral radius are governed by the maximal degree of the vertices, which is much larger for random triangulations (see Ref.\,\cite{Krueger_2015}).
For the \erdoes model one gets a limit of $21.1 \pm 0.2$ and power law exponent $-0.259 \pm 0.006$, for the Newman-Watts model the limit is $15.46 \pm 0.09$ with power law exponent $-0.337 \pm 0.007$.
For the \barabasi model the spectral radius grows much more quickly, because the preferential attachment and the resulting power law degree distribution lead to nodes with much higher degrees than in random triangulations and the other random graphs.
Since in Eq.~\eqref{eq:bounds_spectral_radius} the maximal vertex degree is a lower bound for the spectral radius this implies a higher spectral radius for the \barabasi model.

The parameters obtained by the fits of the spectral radius and the algebraic connectivity are summarised in Tab.~\ref{tab:random_scaling}.

\paragraph{Inverse participation ratio and localization}
\begin{figure}
  \includegraphics{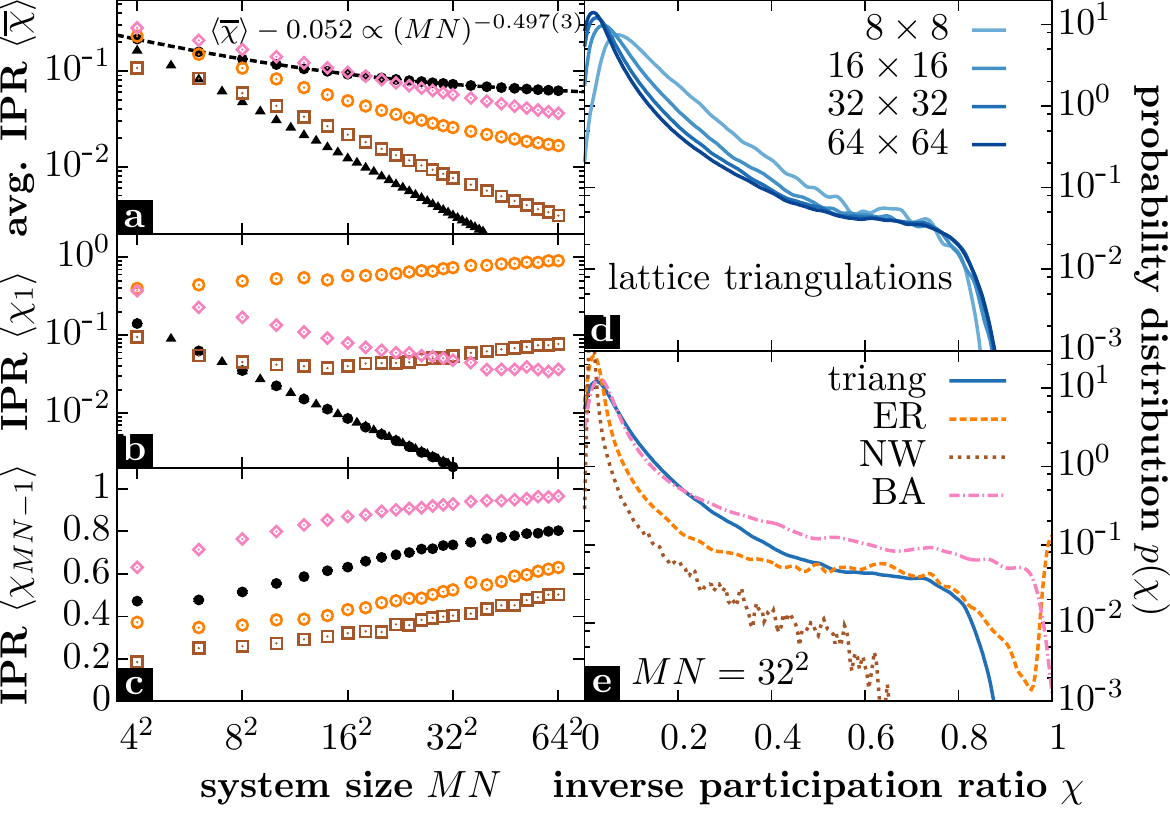}
  \caption{\label{fig:inverse_participation_ratio_random}(Color online) Scaling of the inverse participation ratio (IPR) of the Laplacian spectrum of random triangulations. Expectation values of the average IPR $\langle \overline \chi \rangle$ (a), the IPR $\langle \chi_1 \rangle$ of the algebraic connectivity $\lambda_{1}$ (b) and the IPR of $\langle \chi_{MN-1} \rangle$  of the spectral radius $\lambda_{MN - 1}$ of random triangulations (\protect\includegraphics{point_triangulation.pdf}), maximal ordered triangulations (\protect\includegraphics{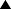}), \erdoes (\protect\includegraphics{point_erdoes_reyni.pdf}), Newman-Watts (\protect\includegraphics{point_newman_watts.pdf}) and \barabasi (\protect\includegraphics{point_barabasi_albert.pdf}) random graphs for different number of vertices $MN$. The displayed lines are power law fits for the numerical data. Probability density funtion $p(\chi)$ of the IPR for random triangulations for different system sizes (d) and for the different random graphs for $32^2$ vertices (e).
  }
\end{figure}

In this section the inverse participation ratio (IPR) \eqref{eq:ipr} of the Laplacian spectrum of random triangulations is examined in terms of the system size.
In Fig.~\ref{fig:inverse_participation_ratio_random} the average IPR $\langle \overline \chi \rangle$, as well as the IPR of the algebraic connectivity $\langle \chi_1 \rangle$ and the spectral radius $\langle \chi_{MN-1} \rangle$ are displayed and compared with the common random graph models.

For random triangulations the average IPR $\langle \overline\chi \rangle$ decreases with a power law 
\begin{equation*}
  \langle \overline\chi \rangle \approx (0.671 \pm 0.008)\cdot (MN)^{-0.512 \pm 0.004} + (0.05334 \pm 0.0005)
\end{equation*}
for increasing system size (similar to the other considered random graph models), the exponent $-0.512 \pm 0.004$ being smaller than the analyical calculated exponent $-1$ for periodic maximal ordered triangulations.
Since the data obtained for random lattice triangulations does not allow for a power-law fit without constant offset, one can conclude that for random triangulations there is a non-vanishing limit $(0.05334 \pm 0.0005)$ for infinite system sizes, in contrast to Newman-Watts and \barabasi random graphs (compare Tab.~\ref{tab:random_scaling}, for \erdoes the situation is ambiguous, both possible fits are done there).
This implies that on average the localization of eigenvectors for random triangulations is higher than for comparable random graphs.

The IPR $\langle \chi_1 \rangle$ of the algebraic connectivity random lattice triangulations is approximatly equal to that of the maximal ordered triangulation and decreases with a power law $\propto x^{-0.896 \pm 0.006}$, since the algebraic connectivity is determined by the vertices with low degree (which are the vertices at the corner of the lattice), and the degree is likely to be unchanged for the random triangulations.
One can find a similar decrease for \barabasi{}, but not for \erdoes or Newman-Watts random graphs (which converge towards a finite value for increasing system size).
The IPR of the spectral radius converges to a value above 0.8, with a similar functional dependency as the other random graph models.

The probability distribution function of the IPRs for random triangulations is comparable to the one of random graphs, there are differences only in the probability of the largest IPRs, which correspond to the strongest localization.

\begin{table}
  \caption{\label{tab:random_scaling}Scaling of algebraic connectivity, spectral radius and average inverse participation ratio with the system size. All values with reduced $\chi^2$ given are fitted using $\lambda = c \pm a\cdot (MN)^b$ or $\lambda = a\cdot (MN)^b$, for the algebraic connectivity of the Newman-Watts and the \barabasi graph the value of the highest system size was taken. The values for the periodic triangular lattice were taken from Sec.~\ref{sec:analytical_solution}.}
  \begin{tabular}{cccc}
    quantity & graph & scaling behavior & reduced $\chi^2$ \\
    \hline\noalign{\smallskip}
    \multirow{5}{*}{$\langle \lambda_1 \rangle$} & per. triangular lat. & $12 \pi (MN)^{-1}$ & - \\
    & rnd. triangs. & $0 + (10.7 \pm 0.2)\cdot (MN)^{-0.949 \pm 0.003}$ & $2.5 \cdot 10^{-8}$\\
    & \erdoes & $\rightarrow 0$ & - \\
    & Newman-Watts & $0.541 \pm 0.003$ & - \\
    & \barabasi & $1.235 \pm 0.001$ & - \\
    \hline\noalign{\smallskip}
    \multirow{5}{*}{$\langle \lambda_{MN-1} \rangle$} & per. triangular lat. & 9 & - \\
    & rnd. triangs. & $(27.4 \pm 0.5) - (27.6 \pm 0.3)\cdot (MN)^{-0.147 \pm 0.006}$ & $4.3 \cdot 10^{-3}$ \\
    & \erdoes & $(21.4 \pm 0.2) - (23.6 \pm 0.2)\cdot (MN)^{-0.236 \pm 0.006}$ & $3.5 \cdot 10^{-3}$ \\
    & Newman-Watts & $(16.01 \pm 0.08) - (20.8 \pm 0.2)\cdot (MN)^{-0.301 \pm 0.006}$ & $2.6 \cdot 10^{-3}$ \\
    & \barabasi & $\rightarrow \infty$ & - \\
    \hline\noalign{\smallskip}
    \multirow{6}{*}{$\langle \overline \chi \rangle$} & per. triangular lat. & $(MN)^{-1}$ & - \\
    & rnd. triangs. & $(0.643 \pm 0.009)\cdot (MN)^{-0.497 \pm 0.004} + (0.0522 \pm 0.0002)$ & $2.4 \cdot 10^{-7}$ \\
    & \erdoes & $(1.19 \pm 0.03)\cdot (MN)^{-0.601 \pm 0.007} + (0.0074 \pm 0.0007)$ & $1.7 \cdot 10^{-6}$ \\
    & & $(1.02 \pm 0.03)\cdot (MN)^{-0.539 \pm 0.007}$ & $8.8 \cdot 10^{-6}$ \\
    & Newman-Watts & $(0.59 \pm 0.05)\cdot (MN)^{-0.58 \pm 0.02}$ & $1.6 \cdot 10^{-5}$ \\
    & \barabasi & $(0.820 \pm 0.006)\cdot (MN)^{-0.383 \pm 0.002}$ & $1.5 \cdot 10^{-6}$ \\
    \hline\noalign{\smallskip}
  \end{tabular}
\end{table}

%% file: results_microcanonical.tex
\subsection{Microcanonical ensemble}\label{subsec:results_microcanonical}

In this section we consider triangulations with a fixed energy which corresponds to a microcanonical ensemble and examine the Laplacian spectrum, the algebraic connectivity $\lambda_1$ and the spectral radius $\lambda_{MN-1}$ in terms of the energy for different system sizes.
For each lattice size we measure the energy in units of the average energy $\langle E \rangle_{\mathrm{rnd}}$ of random triangulations on an equal sized lattice to make the results for different lattice sizes comparable.
We will use $\epsilon = E / E_{\mathrm{rnd}}$ to denote this rescaled energy.

Since in general the Pachner moves are not ergodic if restricting to the subset of triangulations with energy $E$, we use the following algorithm for generating sample triangulations with given energy $E$:
\begin{itemize}
  \item Start with a random triangulation with arbitrary energy, generated as described in Sec.\,\ref{subsec:results_random}.
  \item Perform Metropolis Monte Carlo steps \cite{Metropolis_1953} with the acceptance probability
    \begin{equation}\label{eq:search_acceptance}
      A_{\mathrm{Metropolis}} (\mathcal T_1 \rightarrow \mathcal T_2) := \min \left(1, \frac{\exp[-\beta E(\mathcal T_2)]}{\exp[-\beta E(\mathcal T_1)]} \right)
    \end{equation}
    and check after each step whether the obtained triangulation has the desired energy, then stop (the inverse temperature $\beta$ can be tuned to find the desired energy more quickly).
    Note that the actual number of steps necessary to find a suitable triangulation depends on the given energy and cannot be predicted.
  \item If the desired energy was reached, take the triangulation for measuring the observables and perform $1000MN$ steps at $\beta = 0$ to randomize the triangulation and avoid autocorrelations between successive measurements.
    Note that the autocorrelation time is always below $10MN$ as explained in Sec.\,\ref{subsec:results_random}.
\end{itemize}
These steps are repeated until the desired number of sample triangulations with the correct energy are found.
In this paper we use 1000 samples for each energy and system size.

For relative energies $\epsilon \gtrsim 2$ one must use negative inverse temperatures in Eq.~\eqref{eq:search_acceptance}, but there are many local minima where the simulation can get stuck in, which make searching specific energies difficult.
In these situations we use the acceptance probability
\begin{equation*}
  A_{\mathrm{flat}} (\mathcal T_1 \rightarrow \mathcal T_2) := \min \left(1, \exp\left( H[ E(\mathcal T_1)] - H[ E(\mathcal T_2)] \right) \right),
\end{equation*}
for locating triangulations with the proper energy.
This acceptance probability weights every energy level $E$ with the inverse exponential of the number $H(E)$ the simulation has visited the energy level before.
Intuitively this means the longer the system stays at a certain energy, the larger is the probability for leaving it, which avoids getting stuck in local energy minima.
Basically this procedure is the Wang-Landau algorithm \cite{Wang_2001,Wang_2001b} that will be explained in detail in the following Sec.\,\ref{subsec:results_canonical}, using only a single modification factor.

\begin{figure}
  \includegraphics{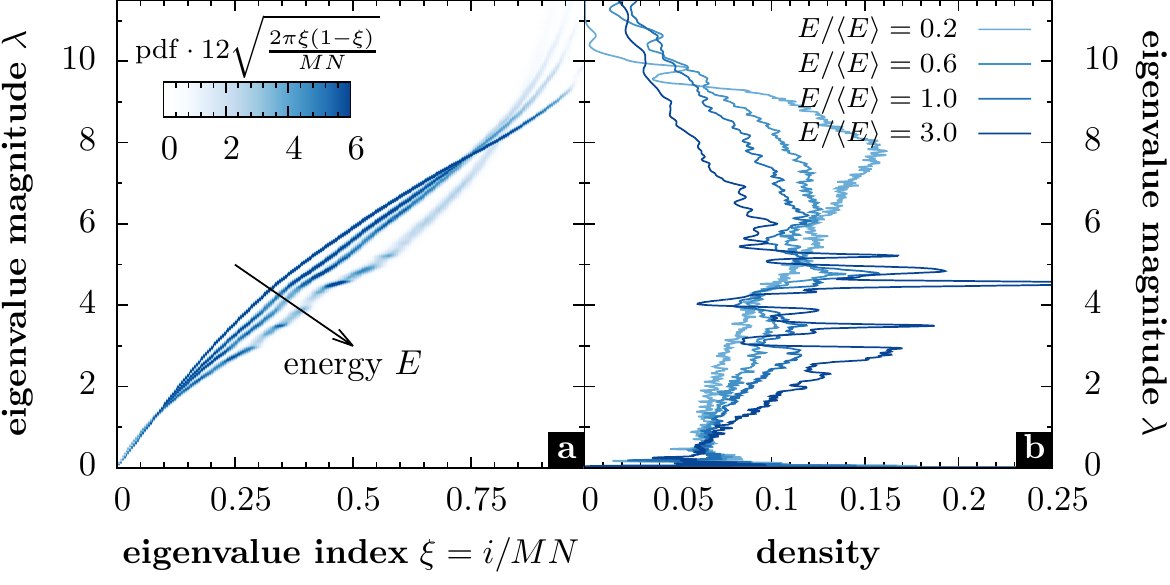}
  \caption{\label{fig:laplacian_spectrum_microcanonical_heatmap}(Color online) Laplacian spectrum of a microcanonical ensemble of $16\times 16$ triangulations. The color code (a) on the left shows the index-resolved probability density function (PDF) in terms of the index of the eigenvalue and the magnitude of the eigenvalues for normalized energies $E / \langle E \rangle_{\mathrm{rnd}} = 0.2, 0.6, 1.0, 3.0$, with $\langle E \rangle_{\mathrm{rnd}} \approx 1000$. The index-summed PDF (b) is also displayed for the same normalized energies.}
\end{figure}

\paragraph{Laplacian spectrum}
The microcanonical average of the Laplacian spectrum probability distribution function (PDF) for different values of the relative energy $\epsilon = E / \langle E \rangle_{\mathrm{rnd}}$ is plotted in Fig.~\ref{fig:laplacian_spectrum_microcanonical_heatmap}. 
With increasing energy the eigenvalues $\lambda_i$ become smaller for relative index $i/MN \lesssim 0.75$ and bigger for relative index $i/MN \gtrsim 0.75$, at index $i / MN \approx 0.75$ the eigenvalues do not change in a significant way.
As expected the index-summed PDF looks similar to the ground state for $\epsilon \gtrsim 0$ with a broad and dominating peak around the eigenvalue magnitude of $\lambda \approx 8$ and a smooth decrease for $\lambda < 6$, and for $\epsilon \approx 1$ it is comparable with the spectrum of the random triangulation with small and narrow peaks emerging.
For relative energies $\epsilon > 2.0$ narrow peaks emerge in the eigenvalue range $2 < \lambda < 6$, which can be seen as nearly horizontal lines in the index-resolved PDF.

\begin{figure}
  \includegraphics{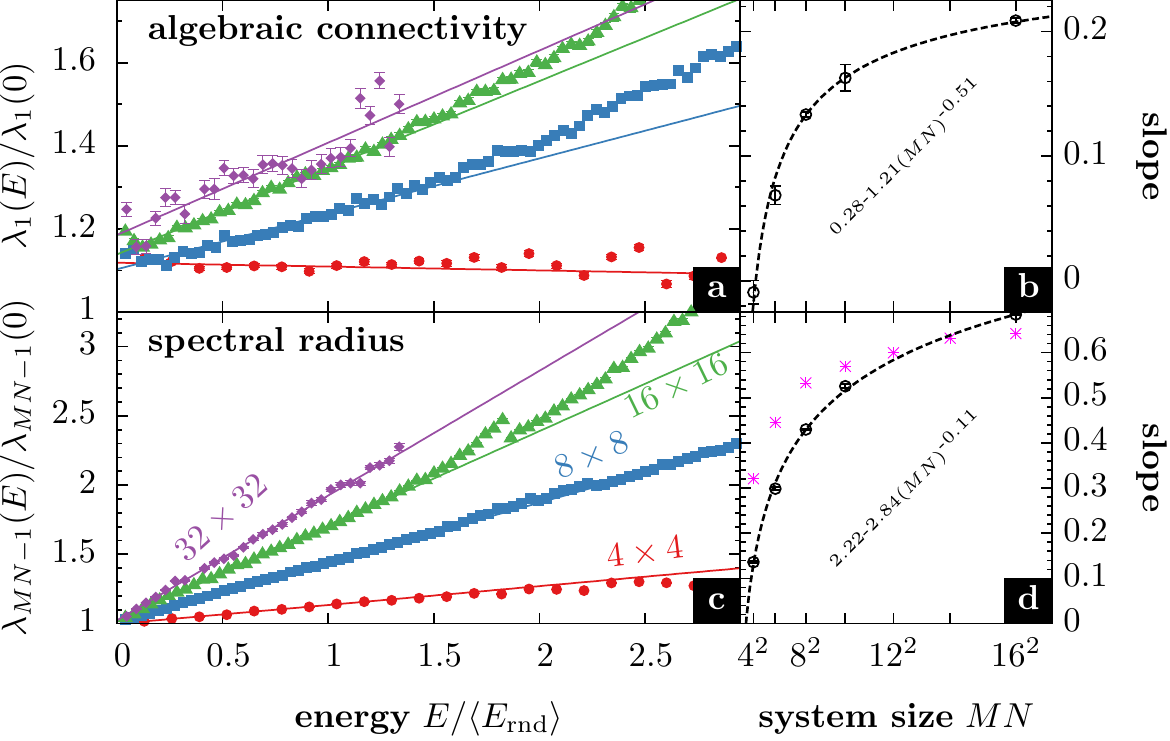}
  \caption{\label{fig:spectrum_observables_microcanonical}(Color online) Dependence of the algebraic connectivity (a) and the spectral radius (c) on the energy of the triangulation normalized by the energy of the random triangulation for $4\times4$-triangulations (\protect\includegraphics{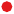}), $8\times 8$-triangulations (\protect\includegraphics{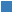}), $16\times 16$-triangulations (\protect\includegraphics{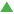}) and $32\times 32$-triangulations (\protect\includegraphics{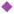}). The solid lines are fits for $E/\langle E_{\mathrm{rnd}} \rangle \in [0,1.5]$. The eigenvalues are normalized by their values at ground states of the respective system size.  (b,d): Gradients of the fits are plotted against the lattice size of the triangulations, the purple points are the approximation for the spectral radius using first order perturbation theory with periodic ground state from Eq.~\eqref{eq:sr_approximation_microcanonical_gradient}.}
\end{figure}

\paragraph{Algebraic connectivity and spectral radius}
Fig.~\ref{fig:spectrum_observables_microcanonical} shows the algebraic connectivity $\lambda_1$, which is the second-smallest Laplacian eigenvalue, and the spectral radius $\lambda_{MN - 1}$, which is the biggest Laplacian eigenvalue, for microcanonical triangulations in terms of the normalized energy for different lattice sizes.
For energies around the average random energy and smaller ($\epsilon = E/\langle E_{\mathrm{rnd}} \rangle \leq 1.5$) both eigenvalues depend linearly on the energy of the triangulation
\begin{equation*}
  \frac{\lambda(E)}{\lambda(0)} \approx m(M,N) \cdot \frac{E}{\langle E\rangle_{\mathrm{rnd}}} + t(M,N)
\end{equation*}
where slope $m(M,N)$ increases with growing system size $M,N$. 
The linear dependence of the eigenvalues is more clearly for the spectral radius than for the algebraic connectivity, and so the fits have smaller errors. 

The slope $m(M,N)$ can be fitted for the algebraic connectivity and the spectral radius by
\begin{equation*}
  \begin{split}
    m_{\mathrm{ac}}(M,N) &= (0.28 \pm 0.02) - (1.21 \pm 0.26)\cdot(MN)^{-0.51\pm 0.09} \\
    m_{\mathrm{sr}}(M,N) &= (2.22 \pm 0.32) - (2.84 \pm 0.27)\cdot(MN)^{-0.11\pm 0.02}
  \end{split}
\end{equation*} 

From Eq.~\eqref{eq:perturbation_theory_spectral_radius} one knows how the spectral radius increases in first order perturbation theory if one performs one step.
One can derive an approximation for the change of the spectral radius in terms of the relative energy $\epsilon = E / \langle E\rangle_{\mathrm{rnd}}$ by assuming that doing several steps at diagonals that are well separated the spectral radius is linear in the number of steps $f$:
\begin{equation*}
  \lambda_{N^2 - 1} \approx 9 + \frac{6f}{N^2} = 9 + \frac{6E}{4N^2} = 9 + \frac{6\langle E\rangle_{\mathrm{rnd}}}{4N^2} \cdot \frac{E}{E\rangle_{\mathrm{rnd}}}
\end{equation*}
Here we used that each step increases the energy by $4$ and so $E = 4f$.
So by using the results for $\langle E\rangle_{\mathrm{rnd}} / N^2$ from \cite{Krueger_2015}, we can find the following relation for the spectral radius:
\begin{equation}\label{eq:sr_approximation_microcanonical_gradient}
  \frac{\lambda_{MN - 1}(\epsilon)}{\lambda_{MN - 1}(0)} = 1 +\frac{1}{6} \cdot \frac{\langle E\rangle_{\mathrm{rnd}}}{MN} \cdot \epsilon
\end{equation}
Comparing Eq.~\ref{eq:sr_approximation_microcanonical_gradient} with the fitted values in Fig.~\ref{fig:spectrum_observables_microcanonical}d one finds a good agreement of the two values for the slope.

%% file: results_canonical.tex
\subsection{Canonical ensemble}\label{subsec:results_canonical}

The results for the microcanonical ensemble of triangulations as calculated in the previous section depend strongly on the choice of the energy function.
In this section we present some results for canonical averages of observables in lattice triangulations.
These have in our opinion the advantage that the results should not change qualitatively if one changes the energy function quantitatively.
Especially the canonical average infinite temperature ($\beta = 0$, which is equivalent to the ensemble of random triangulations) is independent of the choice of the energy function, additionally the limits for $\beta \rightarrow \pm \infty$ should agree qualitatively.

One can use the standard Metropolis Monte-Carlo simulations \cite{Metropolis_1953} for calculating canonical ensemble averages for triangulations by using the acceptance probability \eqref{eq:search_acceptance}, but there is a problem occurring in our setup if trying to access the regime of negative temperatures, which can also be identified with negative coupling and positive temperatures \cite{Knauf_2015,Krueger_2015}:
For high energies there are triangulations that are local minima in the energy landscape, i.e. all triangulations that are connected with these triangulations have a higher energy, so for all possible steps the Metropolis acceptance probabilities \eqref{eq:search_acceptance} are small.
For $\beta < 0$ small enough the algorithm cannot leave these states in the available computation time, although of course there are other states that contribute to the canonical expectation values.
So the Metropolis results are wrong and the system is not computationally ergodic anymore.
Using a parallel tempering approach \cite{Earl_2005} also fails since there is a quasi phase transition for small negative $\beta$ due to the large free energy barrier \cite{Krueger_2015}.

For negative temperatures there can also be a problem with increasing autocorrelation time, as considered numerically in \cite{Krueger_2015}.
A similar problem was considered analytically for lattice triangulations in Refs.\,\cite{Caputo_2015,Stauffer_2015}, where Glauber dynamics was used, which is basically the Metropolis algorithm with a slightly different choice of the acceptance probabilities.
Therein, the authors used an energy function that measures the sum of the edge lengths that qualitatively agrees with the energy function \eqref{eq:energy}.
Instead of using the inverse temperature as control parameter, the coupling constant of the energy was modified and allows to create ordered and disordered lattice triangulations.
It was shown in Ref.\,\cite{Caputo_2015} that the mixing time, which is the time until the Markov chain is near to statistical equilibrium, scales exponentially with the system size for the analog of $\beta < 0$, and polynomial for the analog of a sufficiently high $\beta > \beta_0 > 0$, furthermore it was conjectured that the mixing time is polynomial for all $\beta > 0$.
For random triangulations (which is equal to $\beta = 0$) no result could be obtained.
Although these results are for the mixing and not for the autocorrelation time, one can conjecture a strong relation between both.

\begin{figure}
  \centering
  \includegraphics{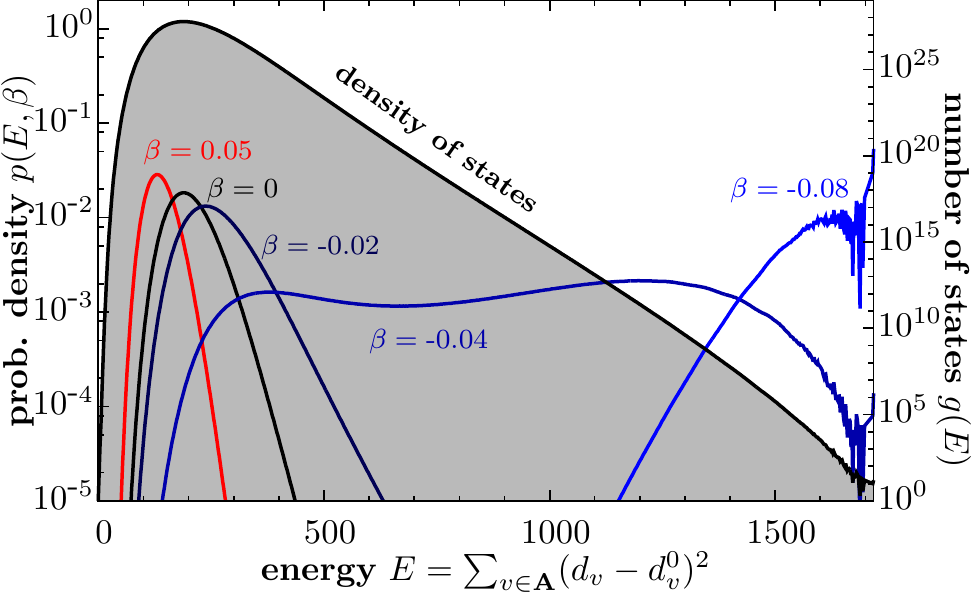}
  \caption{\label{fig:dos_energy_histogram}(Color online) Density of states (black filled curve) for $8\times 8$-triangulations and the probability $P(E)$ (colored lines) for the triangulation to have energy $E$ in the canonical ensemble for different values of the inverse temperature $\beta$.}
\end{figure}

A solution to both problems can be the usage of flat histogram methods as entropic sampling \cite{Lee_1993}, or our method of choice, the Wang-Landau algorithm \cite{Wang_2001,Wang_2001b}:
Let $g(E)$ be the density of states (DOS) of the system, which is the (normalized) number of states with energy $E$.
Instead of sampling according to the Boltzmann distribution $P(\mathcal T) \propto \exp(-\beta E(\mathcal T))$, the Wang-Landau algorithm samples according to the inverse DOS
\begin{equation}\label{eq:wang_landau_acceptance}
  P_{\mathrm{WL}}(\mathcal T) \propto \frac{1}{g\left( E(\mathcal T) \right)} \Rightarrow P_{\mathrm{WL}}(\mathcal T_1 \rightarrow \mathcal T_2) := \min \left(1, \frac{g(E(\mathcal T_1))}{g(E(\mathcal T_2))} \right)
\end{equation}
so each possible value of the energy will be sampled equally often (hence one speaks of \emph{flat-histogram} sampling). 
Of course in most problems the DOS is a prior unknown, such as for lattice triangulations.
The Wang-Landau algorithm starts with an initial guess for the DOS $g_0(E)$ and gradually improves the estimation $g_i(E)$ by increasing the value of the estimation at energy levels that are visited in the simulation by $g_i^\prime(E) = g_i(E) \cdot f_i$.
The modification factor $f_i$ is decreased throughout the simulation whenever the incidence histogram $H(E)$ (which counts the number of visits to the energy level $E$ since the last decrease of the modification factor) is flat.
The estimations then approach the actual DOS $g(E)$ (see \cite{Wang_2001,Wang_2001b} for a detailed discussion of the algorithm, \cite{Knauf_2015} for the numerical calculations of $g(E)$ for 2d lattice triangulations and Fig.~\ref{fig:dos_energy_histogram} for the DOS of $8\times 8$ lattice triangulations).
For our simulations we use an initial modification factor of $f_{\mathrm{initial}} = \exp(1)$ and decrease it to $f_{\mathrm{final}} = \exp(10^{-8})$ using $f_{i+1} = 0.9 \cdot f_i$.
The decrease of the modification factor is chosen carefully compared to $f_{i+1} = 0.5 \cdot f_i$ proposed in \cite{Wang_2001,Wang_2001b} to reduce possible statistical errors.
The incidence histogram $H(E)$  is considered to be flat if $\mathrm{min} H(E) \geq 0.8 \cdot \mathrm{avg} H(E)$.
One simulation for the largest considered system size, $10 \times 10$ lattice triangulations, took around 4 to 5 days of computing time on a single core, and a total of $8.1 \cdot 10^{10}$ attempted steps, which correspond to $3.4 \cdot 10^{8}$ attempted steps per flippable edge in the triangulation.
We performed 10 independent Wang-Landau simulations for system sizes smaller or equal than $8 \times 8$, and 5 independent simulations for the larger system sizes.
The relative statistical error of $g(E)$ is below 0.02 in the former and below 0.03 in the latter case for all energies $E$.

After estimating $g(E)$ of the density of states, canonical averages of an observable $\mathcal O(\mathcal T) = \mathcal O(E(\mathcal T))$ that only depends on the energy can be calculated using
\begin{equation}\label{eq:wl_observable_energy_average}
  \langle \mathcal O \rangle (\beta) = \frac{\sum_E \mathcal O(E) g(E) \exp(-\beta E)}{\sum_E g(E) \exp(-\beta E)}
\end{equation}
This means that the probability $P(E) \propto g(E) \exp(-\beta E)$ to find the system at energy $E$ at inverse temperature $\beta$ weights the observables at the respective energy.
In Fig.~\ref{fig:dos_energy_histogram} the probability $P(E)$ is plotted for different values of the inverse temperature $\beta$, as well as the density of states $g(E)$.

If the desired observable does not depend only on the energy (as the spectrum of the triangulation), but on some other details of the state, one cannot apply Eq.~\eqref{eq:wl_observable_energy_average} and has the following two options: 
First one can extend the simulation and perform a sampling of the extended density of states $g(E,\mathcal O)$ \cite{Kumar_2013}. 
In many cases this is not applicable, such as in lattice triangulations, since the resulting phase space is more complicated and the sampling of the extended density of states needs too much computation time.
The second option is to calculate first the density of states $g(E)$ using the standard Wang-Landau algorithm and then to use the obtained estimation of $g(E)$ to do a flat-histogram sampling with weights \eqref{eq:wang_landau_acceptance} and to record the combined (normalized) histogram $H(E,A)$ which counts the occurrence of observable outcomes $A$ at certain energies $E$ \cite{Gervais_2009,Wuest_2009,Wuest_2012,Kumar_2013}.
The canonical expectation value of the observable can then be calculated using
\begin{equation}\label{eq:wl_observable_other_average}
  \langle \mathcal O \rangle (\beta) 
  = \frac{\sum_E \sum_{\mathcal O} \mathcal O \cdot H(E,\mathcal O) \cdot g(E) \exp(-\beta E)}{\sum_E g(E) \exp(-\beta E)} 
  = \frac{\sum_E \langle \mathcal O \rangle_{\mathrm{mc}} \cdot g(E) \exp(-\beta E)}{\sum_E g(E) \exp(-\beta E)}
\end{equation}
which means that one weights the microcanonical expectation value with the Boltzmann factor.

For two-dimensional unimodular lattice triangulations the density of states can only be calculated for all possible energies up to $11\times 11$ triangulations using the Wang-Landau algorithm, but it is possible to calculate the density of states for energies $E \leq E_c$ smaller than a cutoff $E_c$ for up to $25 \times 25$ triangulations \cite{Knauf_2015}.
This density of states with cutoff can only be used for calculating canonical expectation values for $\beta > \beta_c$, where $\beta_c$ is a cutoff in the inverse temperatures, which is negative for the DOS calculated in \cite{Knauf_2015}.

One has to keep in mind that the regime of negative couplings ($\beta < 0$) is not well defined in the thermodynamic limit (infinite system sizes) since the new specific ground-state energy (which is the negative maximal energy of the positive-coupling model) is not bounded from below \cite{Krueger_2015}.
So one can consider only finite system sizes for $\beta < 0$.

\begin{figure}
  \includegraphics{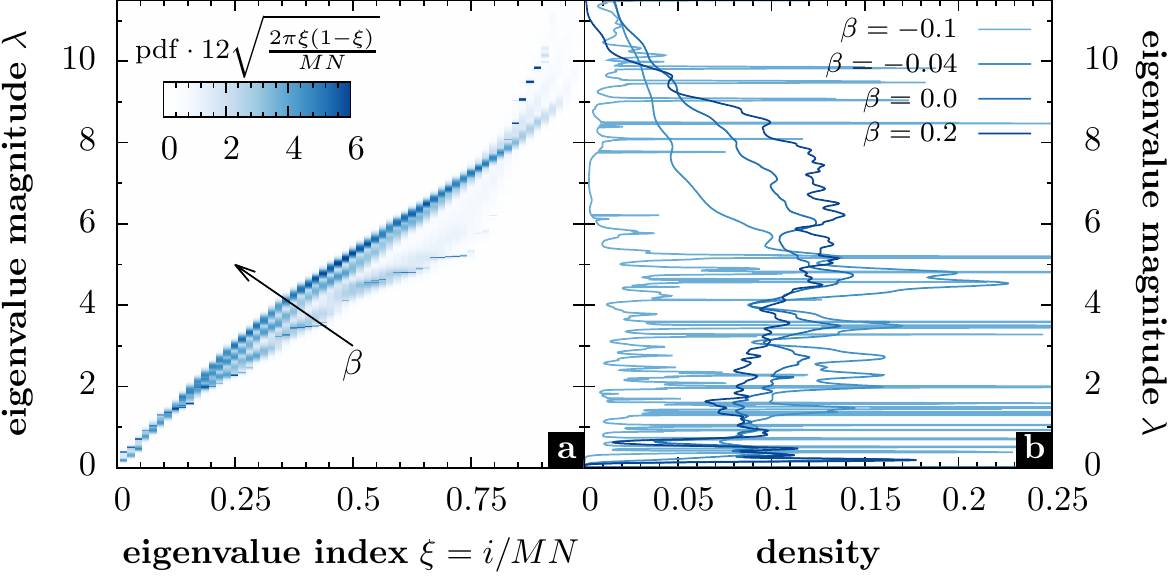}
  \caption{\label{fig:laplacian_spectrum_canonical_heatmap}(Color online) Laplacian spectrum of a canonical ensemble of $8\times 8$ triangulations. The color code (a) left shows the index-resolved probability density function (PDF) in terms of the index of the eigenvalue and the magnitude of the eigenvalues for inverse temperatures $\beta = -0.1, -0.04, 0, 0.2$. The index-summed PDF (b) is also displayed for the different inverse temperatures.}
\end{figure}

\paragraph{Laplacian spectrum}
In Fig.~\ref{fig:laplacian_spectrum_canonical_heatmap} the results for the canonical averages of the Laplacian spectrum of $8\times 8$-triangulations are displayed. 
The indexed-resolved spectrum shows similar behavior as for the different energies in the microcanonical ensemble.
For finite positive temperatures ($\beta = 0.1$) the spectrum approaches the Laplacian spectrum of the ground state.
For finite negative temperatures/couplings ($\beta = -0.04$) the spectrum gets more irregular, which can be seen in the rich peak structure of the summed probability density functions.
This means that the peaked eigenvalues occur very often in the triangulations that contribute to the ensemble average at these temperature with height weight.

\begin{figure}
  \includegraphics{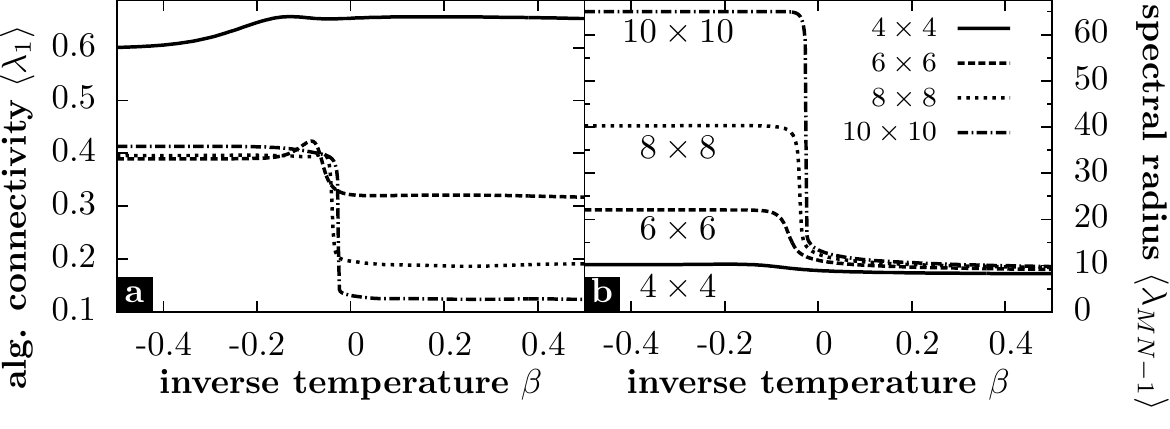}
  \caption{\label{fig:spectrum_observables_canonical}Canonical averages of the algebraic connectivity (a) and spectral radius (b) in terms of the inverse temperature $\beta$ for different lattice sizes, calculated using the Wang-Landau algorithm. The algebraic connectivity is independent of the system size for negative temperatures, whereas the spectral radius is independent of the system size for positive temperatures.}
\end{figure}

\paragraph{Algebraic connectivity and spectral radius}
The results for the canonical expectation values of algebraic connectivity $\lambda_1$ and the spectral radius $\lambda_{MN - 1}$ are displayed in Fig.~\ref{fig:spectrum_observables_canonical} for lattice size $8\times 8$.

Both eigenvalues show the same step-like behavior as the energy, the clustering coefficient and the shortest path length at a negative quasi-critical temperature \cite{Krueger_2015}.
The algebraic connectivity is almost independent of the lattice size for $\beta \rightarrow -\infty$ and decreases for larger lattices in the limit $\beta \rightarrow \infty$, while the spectral radius is approximately independent of the lattice size in the limit $\beta \rightarrow \infty$ and increases for larger lattices in the limit $\beta \rightarrow -\infty$.
The latter behavior can be understood completely by considering the bounds \eqref{eq:bounds_spectral_radius} of the spectral radius.
For $\beta \rightarrow \infty$ the triangulation with the biggest weight are the triangulations near the ground state, where most vertex degrees are $6$ independently of the size, which bounds the spectral radius between $6$ and $12$.
For $\beta \rightarrow -\infty$ the ensemble average is dominated by the triangulation with high energy, which are the triangulations where one edge is connected with almost all possible edges.
Since the degree of this maximal connected vertex, which is the lower bound for the spectral radius, is increasing with the size of the lattice, also the spectral radius must increase.

For a possible application of the canonical ensemble of lattice triangulations one can consider the relation of the algebraic connectivity and the spectral radius with the synchronization time and the return probability \cite{Almendral_2007,Muelken_2011}. 
Here the canonical ensemble with the inverse temperature as control parameter can be used for continuously fine-tuning these quantities on lattice triangulation networks.
For negative inverse temperatures (small-world behavior of the triangulation graph ensemble) with high algebraic connectivity and spectral radius one finds a short synchronization time and a fast decrease of the return probability, which implies a fast delocalization.
For growing inverse temperatures (approaching the large-world behavior of the triangulation graph ensemble) the synchronization time decreases and the delocalization is slower than in the previous region.

\paragraph{Inverse participation ratio}

\begin{figure}
  \includegraphics{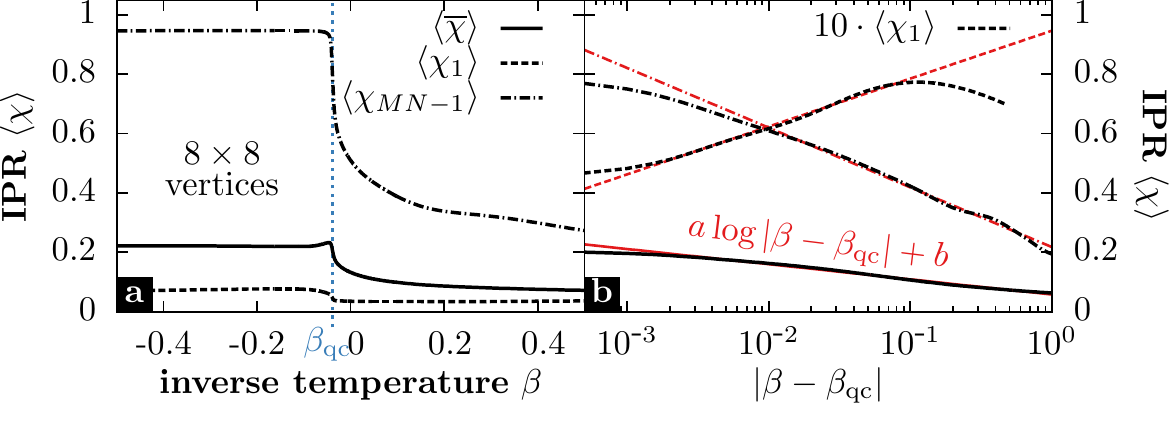}
  \caption{\label{fig:inverse_participation_ratio_canonical}(Color online) Canonical averages of the inverse participation ratio for $8 \times 8$ lattices. (a) Expectation values of the average IPR (black, solid line), the IPR of the algebraic radius (black, dashed line) and of the spectral radius (black, dash-dotted line) in terms of the inverse temperature. (b) The same expectation values in terms of the (logarithmically plotted) distance from a quasi-critical temperature $\beta_{\mathrm{qc}}$, for $\beta > \beta_{\mathrm{qc}}$ ($\langle \overline \chi \rangle$, $\langle  \chi_{MN-1} \rangle$) and for $\beta < \beta_{\mathrm{qc}}$ ($\langle  \chi_{1} \rangle$). The thinner, red lines are linear fits with respect to $\log |\beta - \beta_{\mathrm{qc}}|$. Note that the value of $\langle \chi_1 \rangle$ is streched by a factor 10.}
\end{figure}

In Fig.~\ref{fig:inverse_participation_ratio_canonical} the average inverse participation ratio (IPR) and the IPRs of the algebraic connectivity and the spectral radius are plotted in terms of the inverse temperatures.
As expected, all considered IPRs and therewith the localization increases for increasing disorder in the triangulations.
Near the quasi-critical inverse temperature $\beta_{\mathrm{qc}}$ one observes that the IPRs scale with the logarithm $\log |\beta - \beta_{qc}|$ of the reduced inverse temperature.

%% file: conclusions.tex
\section{Conclusions}\label{sec:conclusions}
In this paper we examined the spectral properties of the graph interpretation of two-dimensional unimodular lattice triangulations.
These triangulations show a transition from ordered large-world to disordered small-world behavior.

For the random triangulations we calculated the spectrum of the adjacency and the Laplacian matrix numerically and compared them with common random graph models.
The algebraic connectivity, which is the smallest non-vanishing eigenvalue of the Laplacian matrix, vanished with a power law in terms of the system size, which is similar to the \erdoes graph, but different to the Newman-Watts and \barabasi random graph, which tend to a fixed value for increasing lattice size.
Calculating the average inverse participation ratio we showed that random lattice triangulations on average exhibit stronger localization than comparable random graphs.

We introduced an energy function that corresponds to the order and the disorder of the lattice triangulation to calculate the dependence of the spectral observables on the order of the triangulation in the microcanonical and the canonical ensemble.

For the microcanonical ensemble we find for small energies a linear dependence of the algebraic connectivity and the spectral radius on the energy of the triangulations.
In the case of the spectral radius the linear dependence can be understood analytically using the ground state values, that can be calculated analytically, and perturbation theory.
For the algebraic connectivity perturbation theory fails because the fixed boundary conditions have stronger influence than the perturbation of the flip.

In the canonical ensemble, that was numerically calculated with the Wang-Landau algorithm, we find a crossover between ordered, large-world behavior with a system-size independent spectral radius and a disordered, small-world behavior with a system-size independent algebraic connectivity and stronger localization than in the ordered phase.
This crossover behavior can also be found in other observables as the mean energy, the clustering coefficient and the average shortest path length \cite{Krueger_2015}. 

Our results can be applied to planar real-world graphs where the actual coordinates of the vertices or the distance between vertices becomes important or must be optimised, e.g. in the power grid or in  transport problems. 
Using also the inserting and removing Pachner moves that lead to non-full triangulations the properties of a grand-canonical ensemble of triangulations can be studied in the future.
In order to take into account arbitrary embedded planar networks, one can also examine triangulations of random point sets instead of a square lattice.
It can also be promising to study the localization in greater detail to understand better the underlying mechanism, e.g.\,on which vertices the eigenvectors of the Laplacian matrix are localized.